\documentclass[reqno, a4paper]{amsart}
\usepackage{amsmath}
\usepackage{amssymb}
\usepackage{amsthm}

\usepackage[scale=0.9]{geometry}

\usepackage{natbib}
\usepackage{bibentry} 

\usepackage[english]{babel}
\usepackage[utf8]{inputenc}

\usepackage{subfig}
\usepackage{graphicx}

\usepackage[unicode, breaklinks]{hyperref}

\usepackage{mathbbol}
\usepackage{bm} 
\usepackage{MnSymbol} 
\usepackage{gensymb}
\usepackage{eurosym}

\usepackage{units}
\usepackage{tensor}
\usepackage{accents}

\usepackage{enumitem}

\usepackage{lineno}

\DeclareMathOperator{\divergence}{div}

\newcommand{\exponential}[1]{\ensuremath{{\mathrm e}^{#1}}}

\newcommand{\reference}{\mathrm{ref}}

\newcommand{\bydefinition}{\mathrm{def}}

\newcommand{\diff}{\mathrm{d}}
\newcommand{\Diff}[1][]{\mathrm{D}_{#1}} 

\renewcommand{\vec}[1]{\ensuremath{\mathbf{#1}}}

\makeatletter
\@ifpackageloaded{bm}%
{\renewcommand{\vec}[1]{\ensuremath{\bm{#1}}}%
}{%
\relax
}
\makeatother
\newcommand{\tensorq}[1]{\ensuremath{\mathbb{#1}}}      

\newcommand{\inverse}[1]{#1^{-1}}

\newcommand{\cstress}{\tensorq{T}}

\makeatletter
\@ifpackageloaded{bm}%
{%
}{%

}

\@ifpackageloaded{bm}%
{%
 
}{%

}

\@ifpackageloaded{bm}%
{%
}{%

}
\makeatother

\newcommand{\gradsym}{\ensuremath{\tensorq{D}}}

\newcommand{\R}{\ensuremath{{\mathbb R}}}

\newcommand{\Prandtl}{\mathrm{Pr}}

\newcommand{\tenergy}{\ensuremath{e}_{\mathrm{tot}}} 
\newcommand{\ienergy}{\ensuremath{e}} 
\newcommand{\fenergy}{\ensuremath{\psi}} 
\newcommand{\entropy}{\ensuremath{\eta}} 

\newcommand{\temp}{\ensuremath{\theta}} 

\newcommand{\nettenergy}{\ensuremath{E}_{\mathrm{tot}}} 
\newcommand{\netentropy}{\ensuremath{S}} 

\newcommand{\cheatvol}{\ensuremath{c_{\mathrm{V}}}}

\newcommand{\efluxc}{\vec{j}_{e}} 

\newcommand{\entfluxc}{\vec{j}_{\entropy}} 

\newcommand{\entprodc}{\xi} 
\newcommand{\entprodctemp}{\zeta} 

\newcommand{\pd}[2]{\ensuremath{\frac{\partial {#1}}{\partial {#2}}}}
\newcommand{\ppd}[2]{\ensuremath{\frac{\partial^2 {#1}}{\partial {#2^2}}}}
\newcommand{\dd}[2]{\ensuremath{\frac{\diff {#1}}{\diff {#2}}}}

\newcommand{\ddd}[2]{\ensuremath{\frac{\diff^2 {#1}}{\diff {#2}^2}}}

\makeatletter
\@ifpackageloaded{tensor}
{

}{%

}
\makeatother

\makeatletter
\@ifpackageloaded{tensor}
{

}{%

}
\makeatother

\newcommand{\norm}[2][]{\ensuremath{\left\|#2\right\|_{#1}}}
\newcommand{\absnorm}[1]{\ensuremath{\left|#1\right|}}

\makeatletter
\@ifundefined{volume}{%
}%
{%
}
\makeatother

\newcommand{\cvolumee}{\diff \mathrm{v}}
\newcommand{\csurfacee}{\diff \vec{s}}

\makeatletter
\@ifpackageloaded{MnSymbol} 
{
\newcommand{\tensordot}[2]{\ensuremath{#1 \vdotdot #2}} 
}{%
\newcommand{\tensordot}[2]{\ensuremath{#1 : #2}} 
}
\makeatother
\newcommand{\vectordot}[2]{\ensuremath{#1 \bullet #2}}

\newcommand{\sleb}[2]{\ensuremath{L}^{#1} \left(#2 \right)}

\newcommand{\entropyrellp}{{\mathcal S}}

\newcommand{\energyrellp}{{\mathcal E}}

\newcommand{\tempbdr}{\temp_{\mathrm{bdr}}}

\newcommand{\tempinit}{\temp_{\mathrm{init}}}

\newcommand{\netstdenergy}{E_{\mathrm{std}}}

\newcommand{\cheatvolref}{\ensuremath{c_{\mathrm{V}, \reference}}}

\newcommand{\vartemp}{\ensuremath{\vartheta}}

\newcommand{\tempref}{\ensuremath{\temp_{\reference}}}

\newcommand{\vartempref}{\ensuremath{\vartemp_{\reference}}}

\newcommand{\kapparef}{\ensuremath{\kappa_{\reference}}}

\title[Thermodynamics and stability in open systems]{Thermodynamics and stability of non-equilibrium steady states in open systems}

\author{Miroslav Bul\'{\i}\v{c}ek}
\email{mbul8060@karlin.mff.cuni.cz}

\author{Josef M\'alek}
\email{malek@karlin.mff.cuni.cz}

\author{V\'{\i}t Pr\r{u}\v{s}a}
\email{prusv@karlin.mff.cuni.cz}

\address{
Faculty of Mathematics and Physics\\
Charles University\\
Sokolovsk\'a 83\\
Praha 8 -- Karl\'{\i}n\\
CZ 186\;75\\
Czech Republic
}

\date{\today}


\numberwithin{equation}{section}

\begin{document}

\begin{abstract}
  Thermodynamical arguments are known to be useful in the construction of physically motivated Lyapunov functionals for nonlinear stability analysis of spatially homogeneous equilibrium steady states in thermodynamically isolated systems. Unfortunately, the limitation to thermodynamically isolated systems is essential, and standard arguments are not applicable even for some very simple thermodynamically open systems.

On the other hand, the nonlinear stability of thermodynamically open systems is usually investigated using the so-called energy method.  Unfortunately, the designation ``energy method'' is clearly a misnomer. The mathematical quantity that is traditionally referred to as the ``energy'' is by no means linked to the energy in the physical sense of the word. Consequently, it would seem that genuine thermodynamical concepts are of no use in the nonlinear stability analysis of thermodynamically open systems.

We show that this is not true. In particular, we propose a construction that in the case of simple heat conduction problem leads to a physically well-motivated Lyapunov functional, which effectively replaces the artificial Lyapunov functional used in the standard energy method. The proposed construction seems to be general enough to be applied in complex thermomechanical settings.


\end{abstract}

\maketitle

\tableofcontents


\section{Introduction}
\label{sec:introduction}

\subsection{Stability of spatially homogeneous equilibrium states in thermodynamically isolated systems}
\label{sec:stab-spat-homog}
Classical thermodynamics of continuous media can be gainfully exploited in nonlinear stability analysis of thermodynamically isolated systems. The physical concepts of net total energy $\nettenergy$ and the net entropy $\netentropy$ help one to design natural Lyapunov functionals for nonlinear stability analysis of \emph{equlibrium rest states} in \emph{thermodynamically isolated} systems. For example, if one is interested in the stability of \emph{equilibrium rest state} of an incompressible Navier--Stokes fluid in a \emph{thermodynamically isolated} vessel, then one can introduce the functional
\begin{equation}
  \label{eq:1}
  \netentropy - \frac{1}{\tempbdr} \nettenergy,
\end{equation}
where $\tempbdr$ denotes the temperature value at the equilibrium rest state, and this functional turns out to be a natural Lyapunov functional characterising the stability of spatially homogeneous equilibrium rest state.

Functionals of the type~\eqref{eq:1} are of use even for very simple \emph{thermodynamically open} systems, namely for systems wherein the temperature value $\tempbdr$ on the boundary is spatially homogeneous, see the seminal contribution by~\cite{coleman.bd:on} and the comprehensive treatise by~\cite{gurtin.me:thermodynamics}. However, if the temperature value $\tempbdr$ on system boundary varies in space, then the standard construction of Lyapunov functional as introduced by~\cite{coleman.bd:on} is inapplicable. (If $\tempbdr$ is a function of position, then expressions of the type~\eqref{eq:1} do not even define a functional.) This restriction is very limiting, since it prevents one from using the construction in the stability analysis of very simple \emph{thermodynamically open} systems such as heat conduction in a differentially heated rigid body. Consequently, nonlinear stability analysis of spatially \emph{inhomogenous non-equilibrium states in thermodynamically open systems} is beyond the reach of the method.  

\subsection{Stability of spatially inhomogeneous non-equilibrium states in thermodynamically open systems}
\label{sec:stab-spat-inhomog}
A mathematical method referred to as the \emph{energy method} has been developed in order to deal with the nonlinear stability of~\emph{inhomogenous non-equilibrium states in thermodynamically open systems}. The method has been used in numerous works on stability of solutions to systems of nonlinear partial differential equations, see for example~\cite{joseph.dd:stability*1,joseph.dd:stability} and \cite{straughan.b:energy} and references therein, and it became the standard method in the field.

The method originated in hydrodynamic stability problems, see~\cite{reynolds.o:on*1} and~\cite{orr.w:stability}, and it was popularised and further elaborated by~\cite{serrin.j:on}. In the original hydrodynamic stability setting the link between the mathematical technique and its physical underpinning is very clear. In hydrodynamic stability problems, the quantity of interest in the mathematical stability analysis is the square of the \emph{Lebesgue norm} $\norm[\sleb{2}{\Omega}]{\vec{v}}$ of the velocity field~$\vec{v}$. This quantity is tantamount, up to a constant, to the \emph{kinetic energy} of the fluid occupying the domain of interest~$\Omega$. Consequently, the name \emph{energy method} is well justified, and one can happily contemplate the close interplay between physics and mathematics.

\subsection{Problems with the concept of energy method}
\label{sec:probl-with-conc}
If the \emph{energy method} is used in a more complex setting such as the nonlinear stability of thermal convection, the name \emph{energy method} becomes problematic. For example, \cite{straughan.b:energy} in his discussion on stability of thermal convection states that
\begin{quotation}
\noindent  At this point, we consider the simplest, natural ``energy'', formed by adding the kinetic and thermal energies of perturbations, and so define $E(t) = \frac{1}{2} \norm{\vec{u}}^2 + \frac{1}{2} \Prandtl \norm{\temp}^2$.
\end{quotation} 
(Please note the quotation marks.) Similarly, \cite{joseph.dd:stability} in his discussion of nonlinear stability of thermosolutal convection states that 
\begin{quotation}
\noindent Though $\frac{\langle \absnorm{\vec{u}}^2\rangle}{2}$ is proportional to the kinetic energy, the other quadratic integrals $\langle \temp^2\rangle$ and $\langle \gamma^2\rangle$ cannot be called energies in any strict sense.
\end{quotation} 
These authors restrain themselves from unequivocally using the word energy for a very good reason. The volume integrals of the square of temperature field, that is the integrals $\norm{\temp}^2 =_{\bydefinition} \int_{\Omega} \temp^2\, \cvolumee$ and $\langle \temp^2 \rangle =_{\bydefinition} \frac{1}{\absnorm{\Omega}} \int_{\Omega} \temp^2\, \cvolumee$, have no clear physical interpretation. In particular, they do not have the meaning of \emph{thermal energy}. This is in striking contrast with the volume integrals $\norm{\vec{u}}^2 =_{\bydefinition} \int_{\Omega} \absnorm{\vec{u}}^2\, \cvolumee$ and $\langle \vec{u}^2 \rangle =_{\bydefinition} \frac{1}{\absnorm{\Omega}} \int_{\Omega} \absnorm{\vec{u}}^2\, \cvolumee$ of the velocity field $\vec{u}$ that are, up to a constant, identical to the \emph{kinetic energy}.

This shows that the name \emph{energy method} is in these settings inappropriate and misleading, although the \emph{mathematical} results obtained on the basis of the \emph{energy method} are of course valid. The problem is that the quantity referred to as the \emph{energy} is no longer the \emph{energy} in the physical sense of the word, and it seems to be a quantity designed purely artificially on the basis of mathematical convenience. The inappropriate name of the method could be seen as a minor issue. It however indicates a more fundamental problem. The former \emph{clear link between the mathematical method and physics is lost}. 

\subsection{Search for physical background of nonlinear stability analysis in thermodynamically open systems}
\label{sec:phys-underp-energy}
The question is whether the presence of the volume integrals of the square of the temperature field, that is of the Lebesgue norm $\norm[\sleb{2}{\Omega}]{\temp}^2 =_{\bydefinition} \int_{\Omega} \temp^2\, \cvolumee$, can be explained/justified by appealing to some other physical concepts than the energy. If this is not possible, one can ask whether there exists another functional of the temperature field that is physically motivated and that can be effectively used as a Lyapunov functional. Further, using such a functional one should be at least able to reach the same conclusions concerning the stability problem for the non-equilibrium steady state in the given~\emph{thermodynamically open system} as the one that can be obtained by the standard energy method. 

Ideally, the construction of a suitable Lyapunov functional for \emph{thermodynamically open systems} should be from the physical point of view as transparent as in the case of nonlinear stability analysis of spatially homogeneous equilibrium rest states in \emph{thermodynamically closed systems}, see~\cite{coleman.bd:on}. One would like to again see a clear connection between physics and the corresponding mathematical method.

\subsection{Heat conduction as a model problem for nonlinear stability analysis in thermodynamically open systems}
\label{sec:therm-cond-as}
Since the core problem is the use of the standard \emph{energy method} in \emph{heat conduction} problems, we investigate the questions in a simple setting of heat conduction in a rigid body. We propose a procedure that leads to the construction of a \emph{physically well motivated Lyapunov functional that regarding the nonlinear stability analysis of a non-equilibrium steady state effectively replaces the artificial squared Lebesgue norm of temperature field}. 

Using the newly designed Lyapunov functional we recover the standard stability result for the heat conduction problem in a rigid body. (Heat conduction governed by the standard Fourier law.) This is of course not a fundamental result. The main outcome of the presented analysis is different. The construction we propose has serious implications regarding the possibility of systematic construction of Lyapunov functionals for nonlinear stability analysis of more complex thermodynamically open systems. This is documented in Appendix~\ref{sec:stab-analys-steady}, where we use the proposed method in the stability study of a heat conduction problem governed by a nonlinear variant of the standard Fourier law.

\subsection{Implications of the proposed construction of physically motivated Lyapunov functional}
\label{sec:implications}
The physically motivated Lyapunov functional for a non-equilibrium steady state in a \emph{thermodynamically open} system is constructed using the physically motivated Lyapunov functional for the equilibrium rest state in the corresponding \emph{thermodynamically isolated} system. Consequently, the construction can be seen as a proper generalisation of the standard thermodynamical procedure introduced by~\cite{coleman.bd:on}. More importantly, the proposed construction of a physically motivated Lyapunov functional seems to be \emph{general enough to be applied in more complex thermomechanical settings} than heat conduction.  Using the proposed construction one can closely follow the physical background behind the given system of governing equations, hence the construction could provide a tool for nonlinear stability analysis of thermomechanical systems that are currently beyond the reach of the standard \emph{energy method}.

\section{Outline}
\label{sec:outline}
The paper is organised as follows. In Section~\ref{sec:heat-cond-rigid} we introduce the stability problem for the steady solution $\widehat{\temp}$ of the heat conduction equation in a rigid body. (Heat conduction governed by the standard Fourier law.) In Section~\ref{sec:stand-proof-uncond} we recall the standard nonlinear stability analysis based on the \emph{energy method}, and we comment in detail on the abuse of the word \emph{energy} in this setting. At the end of Section~\ref{sec:stand-proof-uncond} we rephrase the nonlinear stability analysis as a problem of the design of a suitable Lyapunov functional, which is in the case of the standard energy method given by the formula\footnote{See Section~\ref{sec:heat-cond-rigid} for a detailed discussion of the notation.} 
\begin{equation}
  \label{eq:2}
  \mathcal{V}_{\mathrm{std}} 
  =_{\bydefinition}
  \int_{\Omega} \rho \cheatvol \left( \widetilde{\temp} \right)^2 \, \cvolumee,
\end{equation}
where $\widetilde{\temp}$ denotes the temperature perturbation.

In Section~\ref{sec:uncond-asympt-stab} we propose a physically well motivated construction of a Lyapunov functional suitable for nonlinear stability analysis. (The functional will be different from the square of the Lebesgue norm of the temperature field, that is, from the functional $\mathcal{V}_{\mathrm{std}}$ used in the \emph{energy method}.) In Section~\ref{sec:basic-facts-from}, we recall basic facts from continuum thermodynamics, and then we use thermodynamical concepts in the nonlinear stability analysis. 

First, see Section~\ref{sec:rest-state-equil}, we focus on the stability of the equilibrium rest state in a \emph{thermodynamically isolated} system. (Heat conduction with zero Neumann boundary condition.) The outlined analysis provides an answer to the question \emph{why} one should consider functional~\eqref{eq:1} as a natural candidate for a Lyapunov functional. In this sense, it is complementary to the analysis by~\cite{coleman.bd:on}, who took the functional of type~\eqref{eq:1} as given, and then showed that it actually \emph{is} a Lyapunov functional. 

Second, see Section~\ref{sec:general-steady-state}, we focus on the stability of a \emph{non-equilibrium steady state} in a \emph{thermodynamically open} system. (Heat conduction with inhomogeneous Dirichlet boundary condition.) We use the previously designed Lyapunov functional for the \emph{thermodynamically isolated} system, and we show how to use this functional in designing a new Lyapunov functional 
\begin{equation}
  \label{eq:3}
  \mathcal{V}_{\mathrm{neq}}
  =
  \int_{\Omega}
  \rho
  \cheatvol
  \widehat{\temp}
  \left[
    \frac{\widetilde{\temp}}{\widehat{\temp}}
    -
    \ln \left( 1 + \frac{\widetilde{\temp}}{\widehat{\temp}} \right) 
  \right]
  \,
  \cvolumee
  ,
\end{equation}
that is suitable for this \emph{thermodynamically open} system. The functional $\mathcal{V}_{\mathrm{neq}}$ is argued to be a physically well-justified counterpart of $\mathcal{V}_{\mathrm{std}}$. The results obtained are discussed in Section~\ref{sec:conclusion}.   

Finally, see~Appendix~\ref{sec:stab-analys-steady}, we document the power of the advocated method in the nonlinear stability analysis of heat conduction in a rigid body whose thermal conductivity is a function of temperature. (Heat conduction governed by a nonlinear variant of Fourier law.) In this case we are again dealing with a thermodynamically open system, but its dynamics is now governed by a \emph{nonlinear} partial differential equation. Using the proposed method, we show that the corresponding steady state is unconditionally asymptotically stable.

\section{Stability of heat conduction in a rigid body}
\label{sec:heat-cond-rigid}

\subsection{Governing equation}
\label{sec:governing-equation}
Let us consider a simple problem of heat conduction in a rigid body that occupies a domain~$\Omega$. The evolution of the temperature field $\temp$ in the domain $\Omega$ is governed by the standard heat conduction equation
\begin{equation}
  \label{eq:4}
  \rho \cheatvol \pd{\temp}{t} = \divergence \left( \kappa \nabla \temp \right),
\end{equation}
where $\cheatvol$ denotes the specific heat capacity at constant volume, $[\cheatvol] = \unitfrac{J}{kg \cdot K}$, $\kappa$ denotes the thermal conductivity, $[\kappa] = \unitfrac{W}{m \cdot K}$, and $\rho$ denotes the density, $[\rho] = \unitfrac{kg}{m^3}$. All material parameters are assumed to be constant and positive. Once the initial and boundary conditions are specified, one can solve the equation, and obtain the solution hereafter denoted as $\widehat{\temp}$. The question is whether the solution is stable with respect to perturbations. 

The most studied case is the stability of the steady solution to~\eqref{eq:4} with a prescribed time-independent temperature value $\tempbdr$ on the boundary. This means that~\eqref{eq:4} is supplemented by the boundary condition
\begin{equation}
  \label{eq:5}
  \left. \temp \right|_{\partial \Omega} = \tempbdr.
\end{equation}
In this case the steady solution $\widehat{\temp}$ solves the boundary-value problem
\begin{subequations}
  \label{eq:6}
  \begin{align}
    \label{eq:7}
    0 &= \divergence \left( \kappa \nabla \temp \right), \\
    \label{eq:8}
    \left. \temp \right|_{\partial \Omega} &= \tempbdr. 
  \end{align}
\end{subequations}
This steady solution is usually called the \emph{equilibrium solution}, although it should be rather referred to as a \emph{non-equilibrium steady state}, see Section~\ref{sec:govern-equat-non}. The notion of \emph{equilibrium} solution should be used only for the spatially homogeneous solution to~\eqref{eq:4} in a thermodynamically \emph{closed} system.

\subsection{Stability of steady solution to the governing equation}
\label{sec:stab-steady-solut}
The \emph{nonlinear stability} of the steady non-equilibrium solution~$\widehat{\temp}$ essentially means that any time-dependent temperature field of the form $\temp = \widehat{\temp} + \widetilde{\temp}$ eventually tends to the steady non-equilibrium solution $\widehat{\temp}$ as time goes to infinity. In other words, if the temperature field
\begin{equation}
  \label{eq:9}
  \temp =_{\bydefinition} \widehat{\temp} + \widetilde{\temp} 
\end{equation}
solves the initial-boundary value problem
\begin{subequations}
  \label{eq:10}
  \begin{align}
    \label{eq:11}
    \rho \cheatvol \pd{\temp}{t} &= \divergence \left( \kappa \nabla \temp \right),  \\
    \label{eq:12}
    \left. \temp \right|_{\partial \Omega} &= \tempbdr, \\
    \label{eq:13}
    \left. \temp \right|_{t=0} &= \tempinit,
  \end{align}
\end{subequations}
with an initial temperature distribution $\tempinit$, then one says that the steady non-equilibrium solution $\widehat{\temp}$ is unconditionally asymptotically stable provided that $\widetilde{\temp} \to 0$ as $t \to \infty$ irrespective of the choice of the initial condition. The convergence $\widetilde{\temp} \to 0$ is typically understood as the convergence in a Lebesgue space norm, which under the assumptions granting the regularity of the solution, implies also the pointwise convergence everywhere in the domain $\Omega$.

The adjective nonlinear means that we are interested in the stability with respect to finite perturbations, and that we are not dealing with the dynamics of the linearised equations in the neighborhood of the steady state as in the standard linearised stability theory, see for example \cite{lin.cc:theory}, \cite{chandrasekhar.s:hydrodynamic}, \cite{yudovich.vi:linearization}, \cite{drazin.pg.reid.wh:hydrodynamic} or~\cite{schmid.pj.henningson.ds:stability}.

\section{Unconditional asymptotic stability of steady non-equilibrium solution -- the standard proof}
\label{sec:stand-proof-uncond}

\subsection{Standard energy method}
\label{sec:energy-method}
The standard \emph{energy method} based proof of unconditionally asymptotic stability of a steady non-equilibrium solution to~\eqref{eq:4} with boundary condition~\eqref{eq:5} proceeds as follows. 

First, one formulates the governing equations for the perturbation $\widetilde{\temp}$. Since $\temp = \widehat{\temp} + \widetilde{\temp}$ solves~\eqref{eq:10} and $\widehat{\temp}$ solves~\eqref{eq:6}, the governing equations for the perturbation $\widetilde{\temp}$ read
\begin{subequations}
  \label{eq:14}
  \begin{align}
    \label{eq:15}
    \rho \cheatvol \pd{\widetilde{\temp}}{t} &= \divergence \left( \kappa \nabla \widetilde{\temp} \right),  \\
    \label{eq:16}
    \left. \widetilde{\temp} \right|_{\partial \Omega} &= 0, \\
    \label{eq:17}
    \left. \widetilde{\temp} \right|_{t=0} &= \tempinit - \widehat{\temp}.
  \end{align}
\end{subequations}

Second, one multiplies the evolution equation~\eqref{eq:15} by $\widetilde{\temp}$, integrates over the domain $\Omega$, and then uses integration by parts in the term $\int_{\Omega} \divergence \left( \kappa \nabla \widetilde{\temp} \right) \widetilde{\temp} \, \cvolumee$. The boundary term in the integration by parts formula vanishes in virtue of the boundary condition~\eqref{eq:16}, and one obtains the equality\footnote{Symbol $\vectordot{\vec{a}}{\vec{b}}$ denotes the standard scalar product of two vectors in $\R^3$.}
\begin{equation}
  \label{eq:18}
  \frac{1}{2}
   \dd{}{t} \int_{\Omega} \rho \cheatvol \left( \widetilde{\temp} \right)^2 \, \cvolumee = - \int_{\Omega}  \kappa \vectordot{\nabla \widetilde{\temp}}{\nabla \widetilde{\temp}} \, \cvolumee.
\end{equation}
This equation is the evolution equation for the quantity 
\begin{equation}
  \label{eq:19}
  \netstdenergy =_{\bydefinition} \frac{1}{2} \int_{\Omega} \rho \cheatvol \left(\widetilde{\temp} \right)^2\, \cvolumee,
\end{equation}
which is commonly referred to as the \emph{energy} of the perturbation~$\widetilde{\temp}$ or the \emph{energy norm} of the perturbation~$\widetilde{\temp}$, see for example~\cite{joseph.dd:stability} or~\cite{straughan.b:energy}. Equation~\eqref{eq:18} shows that the energy $\netstdenergy$ of the perturbation decays in time, $\dd{\netstdenergy}{t} \leq 0$, which essentially finishes the proof of unconditional asymptotic stability of the solution $\widehat{\temp}$. 

Moreover, using the standard Poincar\'e inequality, see for example~\cite{gilbarg.d.trudinger.ns:elliptic} or~\cite{evans.lc:partial}, it is easy to show that the norm of the perturbation decays to zero exponentially fast. A similar argument can be carried out also if the Dirichlet boundary condition~\eqref{eq:5} is replaced by the zero Neumann boundary condition $\left. \vectordot{\nabla \temp}{\vec{n}} \right|_{\partial \Omega} = 0$, where~$\vec{n}$ denotes the outward unit normal to the boundary of the domain $\Omega$, see for example~\cite{stein.j.prusa.v:viscoelastic} for a worked-out example.

\subsection{Remarks on the notion of energy}
\label{sec:remarks-noti-energy}
The standard proof is correct, and gives one a tool to prove the desirable proposition concerning asymptotic stability. However, the terminology \emph{energy norm} or \emph{energy} for the quantity $\netstdenergy$ defined via~\eqref{eq:19} is inappropriate and misleading for several reasons. 

\emph{First}, the quantity $\netstdenergy$ does not even have the physical dimension of \emph{physical} energy. \emph{Second}, even if the physical dimension were corrected by a suitable constant multiplicative factor, the integral~\eqref{eq:19} would be different from the \emph{physical} net total energy of the perturbation. Indeed, the \emph{physical} net total energy is in the present case given by the formula
\begin{equation}
  \label{eq:20}
  \nettenergy
  =_{\bydefinition}
  \int_{\Omega}
  \rho
  \cheatvol
  \temp
  \,
  \cvolumee
  ,
\end{equation}
hence the net total energy of the perturbation $\widetilde{\temp}$ reduces to 
$
\int_{\Omega}
\rho
\cheatvol
\widetilde{\temp}
\,
\cvolumee
$,
which is different from~\eqref{eq:19}. \emph{Third}, the term ``energy'' for the quantity $\netstdenergy$ is used even if one studies the stability of thermodynamically isolated system. However, in such a system the \emph{physical} energy is a quantity that is \emph{constant} in time, and it provides almost no clue concerning the evolution of the perturbation~$\widetilde{\temp}$. In particular, it can not be used for the characterisation of the \emph{decay in time}.

Consequently, the quantity $\netstdenergy$ \emph{should not be referred to as the energy}. (At least when one wishes to understand the term energy as a term that has a physical meaning.) The proper term should be the mathematical one. Quantity $\netstdenergy$ is, up to a constant multiplicative factor, the square of the norm of the perturbed temperature field $\widetilde{\temp}$ in the Lebesgue space~$\sleb{2}{\Omega}$. 

Now one is tempted to claim that the stability problem can not be solved by appealing to some physical concepts. Indeed, since the outlined proof is based on the \emph{mathematical} concept of the norm in a Lebesgue space, one can argue that the true physical quantities such as the net total energy or the net entropy play no substantial role in the stability theory\footnote{Note that the situation is different in hydrodynamic stability theory, see for example~\cite{serrin.j:on}. There the norm of the velocity perturbation in Lebesgue space $\sleb{2}{\Omega}$ is, up to a constant multiplicative factor, tantamount to the physical \emph{net kinetic energy} of the perturbation.}. Consequently, the stability problem seems to be a purely mathematical problem that must be solved only by mathematically motivated manipulations with the governing equations. 

As we shall demonstrate below this is not the case. In fact we show that thermodynamics plays a substantial role in nonlinear stability analysis. Moreover, we show that this is true even \emph{in the case of thermodynamically open systems}.

\subsection{Energy method from the perspective of Lyapunov method}
\label{sec:energy-method-from}

Note that one can rephrase the outlined proof using the concept of Lyapunov functional. The concept was introduced by \cite{lyapunov.am:general} for the analysis of the stability of solutions to ordinary differential equations, see also~\cite{la-salle.j.lefschetz.s:stability}. However, the concept works equally well for analysis of the stability of solutions to partial differential equations, see for example~\cite{yoshizawa.t:stability} and~\cite{henry.d:geometric}. 

Using the concept of Lyapunov functional, one can say that the square of Lebesgue norm $\norm[\sleb{2}{\Omega}]{\cdot}$ of the perturbation $\widetilde{\temp}$ is a natural Lyapunov functional characterising the stability of the equilibrium solution $\widehat{\temp}$. Indeed, the functional 
$
\mathcal{V}_{\mathrm{std}} 
\left(
  \temp
\right)
$
defined as
\begin{equation}
  \label{eq:21}
  \mathcal{V}_{\mathrm{std}} 
  \left(
    \temp
  \right)
  =_{\bydefinition}
  \int_{\Omega} \rho \cheatvol \left( \temp - \widehat{\temp} \right)^2 \, \cvolumee,
\end{equation}
is nonnegative and it vanishes if and only if $\temp = \widehat{\temp}$ in $\Omega$, that is if and only if the steady equilibrium solution is attained. Further, the time derivative of the functional is negative along the trajectories determined by the corresponding governing equation~\eqref{eq:10}. This is easy to see if the temperature field $\temp$ is written in the form $\temp = \widehat{\temp} + \widetilde{\temp}$, which shows that the Lyapunov functional $\mathcal{V}_{\mathrm{std}}$ is in fact identical, up to a constant coefficient, to the ``energy'' $\netstdenergy$ of the perturbation as introduced in~\eqref{eq:19}.

Now the question is the same. Is the choice of Lyapunov functional $\mathcal{V}_{\mathrm{std}}$ motivated by a physical insight or is it just a matter of mathematical convenience? 

\section{Unconditional asymptotic stability -- a proof using concepts from non-equilibrium thermodynamics}
\label{sec:uncond-asympt-stab}

\subsection{Basic facts from thermodynamics of continuous media}
\label{sec:basic-facts-from}
Before proceeding with the thermodynamical analysis, let us recall some basic textbook facts from nonequilibrium continuum thermodynamics\footnote{The formulae below are straightforward generalisations of the standard formulae from classical equilibrium thermodynamics, see for example~\cite{callen.hb:thermodynamics}, to the setting of spatially distributed fields. See for example~\cite{muller.i:thermodynamics} for details.} that are necessary for correct understanding of the physical background of the evolution equation~\eqref{eq:4}.

\subsubsection{Specific Helmholtz free energy, specific entropy, specific internal energy }
\label{sec:spec-helmh-free}
First, if the rigid body of interest has a constant specific heat capacity at constant volume $\cheatvol$, then the body can be characterised by the specific Helmholtz free energy $\fenergy$, $[\fenergy] = \unitfrac{J}{kg}$, in the form
\begin{equation}
  \label{eq:22}
  \fenergy 
  =_{\bydefinition} 
  - 
  \cheatvol \temp \left(\ln \left( \frac{\temp}{\temp_{\reference}} \right) - 1 \right),
\end{equation}
where $\temp_{\reference}$ is a constant reference temperature value. Note that the specification of the Helmholtz free energy in fact determines how the body \emph{stores the energy}, and this piece of information is usually the key starting point for modern theories of constitutive relations in continuum thermodynamics, see for example~\cite{rajagopal.kr.srinivasa.ar:on*7} or~\cite{malek.j.prusa.v:derivation} for details. In particular, formulae for the specific Helmholtz free energy are known for many materials that are far more complex than the rigid heat conducting material, see for example~\cite{dressler.m.edwards.bj.ea:macroscopic} or \cite{hron.j.milos.v.ea:on} for the case of polymeric liquids.

The formula for the specific entropy $\entropy$, $[\entropy] = \unitfrac{J}{kg \cdot K}$, is obtained by differentiating the specific Helmholtz free energy $\fenergy$ with respect to the temperature,
\begin{equation}
  \label{eq:23}
  \entropy = -\pd{\fenergy}{\temp}.
\end{equation}
In particular, for the specific Helmholtz free energy $\fenergy$ in the form~\eqref{eq:22} we get
\begin{equation}
  \label{eq:24}
  \entropy 
  = 
  \cheatvol \ln \left( \frac{\temp}{\temp_{\reference}} \right).
\end{equation}

The specific internal energy $\ienergy$, $[\ienergy] = \unitfrac{J}{kg}$ and the specific Helmholtz free energy $\fenergy$ are related via Legendre transformation $\fenergy = \ienergy - \temp \entropy$. This in our simple case yields
\begin{equation}
  \label{eq:25}
  \ienergy = \cheatvol \temp.
\end{equation}

\subsubsection{Entropy production}
\label{sec:entropy-production}
Second, one needs to characterise the entropy production mechanisms in the body. Again, this piece of information is crucial in modern theory of constitutive relations in continuum thermodynamics, and entropy production mechanisms are known for many complex materials. In the present case, the entropy production is given by the formula $\entprodc = \frac{\entprodctemp}{\temp}$, where
\begin{equation}
  \label{eq:26}
  \entprodctemp =_{\bydefinition} \kappa \frac{\absnorm{\nabla \temp}^2}{\temp}.
\end{equation}
If~\eqref{eq:26} holds, then the energy flux $\efluxc$ in the body is given by the classical Fourier law
\begin{equation}
  \label{eq:27}
  \efluxc = - \kappa \nabla \temp,
\end{equation}
and the entropy flux $\entfluxc$ is given by the standard formula $\entfluxc = \frac{\efluxc}{\temp}$.

\subsubsection{Evolution equations for the total energy, specific internal energy and specific entropy}
\label{sec:evol-equat-entr}
Finally, the generic evolution equations for the specific total energy $\tenergy = \ienergy + \frac{1}{2} \absnorm{\vec{v}}^2$, specific internal energy $\ienergy$ and specific entropy $\entropy$ read, in the absence of external forces and heat sources, as follows
\begin{subequations}
  \label{eq:28}
  \begin{align}
    \label{eq:29}
    \rho \dd{}{t}\left( \ienergy + \frac{1}{2} \absnorm{\vec{v}}^2 \right) &= \divergence \left(\cstress \vec{v} - \efluxc \right), \\
    \label{eq:30}
    \rho \dd{\ienergy}{t} &= \tensordot{\cstress}{\gradsym} - \divergence \efluxc, 
    \\
    \label{eq:31}
    \rho
    \dd{\entropy}{t}
    &=
    \frac{\entprodctemp}{\temp}
    -
    \divergence
    \entfluxc
    ,
  \end{align}
\end{subequations}
see for example~\cite{truesdell.c.noll.w:non-linear}. Here $\cstress$ denotes the Cauchy stress tensor, $\vec{v}$ denotes the velocity field, $\gradsym$ denotes the symmetric part of the velocity gradient, and $\dd{}{t}$ denotes the material time derivative, that is for any quantity $\varphi$ we have $\dd{\varphi}{t} =_{\bydefinition} \pd{\varphi}{t} + \vectordot{\vec{v}}{\nabla \varphi}$. Symbol $\absnorm{\vec{v}}$ denotes the norm induced by the standard scalar product in $\R^3$. In the case of heat conduction in a rigid body one has $\vec{v}=0$, hence $\tensordot{\cstress}{\gradsym} = 0$, and the material time derivative $\dd{}{t}$ coincides with the partial time derivative $\pd{}{t}$. The heat conduction equation~\eqref{eq:4} is then obtained by the substitution of~\eqref{eq:25} and~\eqref{eq:27} into~\eqref{eq:30}.

\subsubsection{Net total energy, net entropy}
\label{sec:net-total-energy}
Having explicit formulae for the specific internal energy $\ienergy$ and the specific entropy $\entropy$, we can explicitly identify the \emph{net total energy} $\nettenergy$ and \emph{net entropy} $\netentropy$ of the body occupying the domain $\Omega$,
\begin{subequations}
  \label{eq:net-total-energy-and-entropy}
  \begin{align}
    \label{eq:32}
    \nettenergy
    &=_{\bydefinition}
    \int_{\Omega}
    \rho
    \left[
      \frac{1}{2} \absnorm{\vec{v}}^2
      +
      \ienergy
    \right]
    \,
    \cvolumee
    ,
    \\
    \label{eq:33}
    \netentropy
    &=_{\bydefinition}
    \int_{\Omega}
    \rho
    \entropy
    \,
    \cvolumee
    .
  \end{align}
\end{subequations}

Note that in the studied case of heat conduction in a rigid body the kinetic energy contribution $\int_{\Omega} \rho  \frac{1}{2}\absnorm{\vec{v}}^2 \, \cvolumee$ in~\eqref{eq:32} vanishes since we consider a fixed rigid body with $\vec{v}=\vec{0}$. Formula~\eqref{eq:32} however holds even for a moving continuous medium and it is written down for the sake of completeness. Since the concepts of the net total energy and net entropy are apparently well defined whenever one has an expression for the specific Helmholtz free energy, we see that these concepts are not exclusively restricted to the studied case of heat conduction in a fixed rigid body.  

\subsubsection{Thermodynamically isolated system}
\label{sec:therm-isol-syst}
Once we have explicit formulae for the \emph{energy flux} and the \emph{entropy flux}, we know what boundary conditions imply that the system of interest is thermodynamically isolated. (Thermodynamically isolated system is a system that is not allowed to exchange any form of energy with its surrounding.) The boundary condition that express the fact that the body is thermodynamically isolated is $\left. \vectordot{\left(\cstress \vec{v} - \efluxc \right)}{\vec{n}} \right|_{\partial \Omega}=0$, which in our setting translates to
\begin{equation}
  \label{eq:34}
  \left. \vectordot{\nabla \temp}{\vec{n}} \right|_{\partial \Omega} = 0,
\end{equation}
where $\vec{n}$ denotes the unit outward normal to $\Omega$. Note that if the body is thermodynamically isolated then~\eqref{eq:29} implies that the net total energy is conserved, $\dd{\nettenergy}{t} = 0$. 

\subsection{Unconditional asymptotic stability of the rest state in a thermodynamically isolated system}
\label{sec:rest-state-equil}
Now we are in the position to exploit thermodynamical concepts in nonlinear stability analysis. First, we investigate the stability of the spatially homogeneous equilibrium rest state in a \emph{thermodynamically isolated} body, and then we proceed with the stability analysis of a steady state in a \emph{thermodynamically open} setting. The stability problem for the spatially homogeneous equilibrium rest state is in fact a very simple problem, but it will motivate the techniques used in a more general setting. In both cases, we show that thermodynamical concepts can be used in a systematic construction of Lyapunov functionals characterising the stability of the corresponding solution.

\subsubsection{Governing equations for the equilibrium rest state}
\label{sec:govern-equat-equil}
The \emph{steady} solution $\widehat{\temp}$ of~\eqref{eq:4} with the boundary condition~\eqref{eq:34}, that is of the system
\begin{subequations}
  \label{eq:35}
  \begin{align}
    \label{eq:36}
    0 &= \divergence \left( \kappa \nabla \temp \right), \\
    \label{eq:37}
    \left. \vectordot{\nabla \temp}{\vec{n}} \right|_{\partial \Omega} &= 0,
  \end{align}
\end{subequations}
is a spatially homogeneous constant temperature field $\widehat{\temp} = \tempbdr$. The value of $\tempbdr$ corresponds to the initial value of the net total energy~$\widehat{\nettenergy}$, that is 
\begin{equation}
  \label{eq:38}
  \tempbdr = \frac{\widehat{\nettenergy}}{\rho \cheatvol \absnorm{\Omega}},
\end{equation}
where $\absnorm{\Omega}$ denotes the volume of the domain occupied by the rigid body.

In other words, the equilibrium rest state temperature distribution in a thermodynamically isolated rigid body is spatially homogeneous. In particular, the temperature value inside the body corresponds to the temperature value on the boundary. Since $\widehat{\temp}$ is a constant, we see that the associated \emph{entropy production} given by~\eqref{eq:26} \emph{vanishes}. This means that the temperature distribution $\widehat{\temp}$ attained at the equilibrium rest state in the thermodynamically isolated body indeed deserves to be referred to as an \emph{equilibrium} temperature distribution. Moreover, the physical notion of equilibrium (zero entropy production) coincides with the dynamical systems theory notion of equilibrium (right-hand side of~\eqref{eq:4} vanishes).

\subsubsection{Governing equations for the perturbation}
\label{sec:govern-equat-pert}
We are interested in the stability of the equilibrium rest state $\widehat{\temp}$, which means that we need to solve the evolution equations
\begin{subequations}
  \label{eq:39}
  \begin{align}
    \label{eq:40}
    \rho \cheatvol \pd{\temp}{t} &= \divergence \left( \kappa \nabla \temp \right),  \\
    \label{eq:41}
    \left. \vectordot{\nabla \temp}{\vec{n}} \right|_{\partial \Omega} &= 0, \\ 
    \label{eq:42}
    \left. \temp \right|_{t=0} &= \tempinit,
  \end{align}
\end{subequations}
and show that $\temp \to \widehat{\temp}$ at $t \to +\infty$ for any initial spatially inhomogeneous temperature field $\tempinit$. The initial temperature field  $\tempinit$ can be arbitrary, but it must satisfy some natural compatibility requirements. First, the initial temperature field $\tempinit$ must be positive at every point of the domain. Second, the initial temperature field must be compatible with the given net total energy $\widehat{\nettenergy}$. (Net total energy must be conserved in thermodynamically isolated systems.) In other words, we require $\nettenergy = \widehat{\nettenergy}$, which reduces to
\begin{equation}
  \label{eq:43}
  \int_{\Omega}
  \rho
  \cheatvol
  \tempinit
  \,
  \cvolumee
  =
  \int_{\Omega}
  \rho
  \cheatvol
  \tempbdr
  \,
  \cvolumee
  .
\end{equation}

\subsubsection{Construction of a physically motivated Lyapunov functional -- an unsuccessful attempt}
\label{sec:constr-phys-motiv}
When investigating the stability of the equilibrium steady state $\widehat{\temp}$, we would like to identify a suitable Lyapunov functional. A natural \emph{physically} motivated candidate for Lyapunov functional is the (negative) net entropy $\netentropy$, since the net entropy $\netentropy$ is in a thermodynamically isolated system a nondecreasing function of time. This is easy to see by integrating~\eqref{eq:31} over the domain $\Omega$, which yields
\begin{equation}
  \label{eq:44}
  \dd{\netentropy}{t} = \int_{\Omega} \frac{\entprodctemp}{\temp} \, \cvolumee \geq 0,
\end{equation}
where the \emph{entropy flux} $\entfluxc$ vanishes in virtue of the boundary condition~\eqref{eq:41}. 

The explicit formula for the net entropy functional $\netentropy$ in our case reads
\begin{equation}
  \label{eq:45}
  \netentropy
  =
  \int_{\Omega}
  \rho
  \cheatvol \ln \left( \frac{\temp}{\temp_{\reference}} \right)
  \,
  \cvolumee
  ,
\end{equation}
where the reference temperature $\temp_{\reference}$ can be, for the sake of convenience, fixed as 
\begin{equation}
  \label{eq:46}
  \temp_{\reference} = \tempbdr = \widehat{\temp}.
\end{equation}
Consequently, we see that $\netentropy(\widehat{\temp} + \widetilde{\temp})$ vanishes provided that $\widetilde{\temp} = 0$, which is a desirable property in the construction of Lyapunov functional. 

However, the \emph{net entropy functional does not provide sufficient information on the spatial distribution of the temperature}. In other words, $\netentropy(\widehat{\temp} + \widetilde{\temp}) = 0$ does not imply $\widetilde{\temp} = 0$, and, much worse, the functional can be both positive or negative depending on the particular choice of $\widetilde{\temp}$. Consequently, the functional does not provide a well defined notion of ``distance'' between the steady equilibrium solution and its perturbation, and it can not be used as a Lyapunov functional.

Does this mean that thermodynamics has nothing to say with respect to the construction of Lyapunov functional? Absolutely not. One has to recall that thermodynamics is based on two concepts -- the \emph{entropy} and the \emph{energy}. One should not be dealt with in the absence of the other. Indeed, we can construct a suitable thermodynamically motivated Lyapunov functional if we use the energy in addition to the entropy. 

\subsubsection{Construction of a physically motivated Lyapunov functional -- a successful attempt}
We will exploit the famous formulation of the first and second law of thermodynamics by \cite{clausius.r:ueber}, namely the following statement
\begin{quotation}
  The energy of the world is constant. The entropy of the world strives to a maximum.
\end{quotation}
In other words, the entropy of a \emph{thermodynamically isolated system} attains \emph{in the long run} the maximal possible value achievable at the given energy level. Note that although the original statement was formulated for spatially homogeneous systems, we can use it with a little modification also for spatially inhomogenenous systems. The only modification is that the energy and the entropy must be understood as the \emph{net total energy} and the \emph{net entropy}.

The maximum net entropy value achievable at the given net total energy level can be determined by solving a constrained maximisation problem. We want to maximise the \emph{net entropy}~\eqref{eq:45} over all possible temperature fields $\temp$ that satisfy~\eqref{eq:41} and that have the \emph{net total energy} $\nettenergy$ equal to the reference net total energy $\widehat{\nettenergy}$. This is easy to do using the Lagrange multiplier technique. The auxiliary functional for the constrained maximisation problem is
\begin{equation}
  \label{eq:47}
    \mathcal{L}_{\Lambda}(\temp) = _{\bydefinition} \netentropy - \Lambda \left( \nettenergy - \widehat{\nettenergy} \right),
\end{equation}
where $\Lambda$ is the Lagrange multiplier. The G\^ateaux derivative\footnote{%
\label{fn:1}
Let us recall that the G\^ateaux derivative $\Diff \mathcal{M} (\vec{x})[\vec{y}]$ of a functional $\mathcal{M}$ at point $\vec{x}$ in the direction $\vec{y}$ is defined as
$
    \Diff \mathcal{M} (\vec{x})[\vec{y}]
    =_{\bydefinition}
    \lim_{s \to 0}
    \frac{
      \mathcal{M} (\vec{x} + s \vec{y}) - \mathcal{M} (\vec{x})
    }
    {
      s
    }
$ 
which is tantamount to 
$
\Diff \mathcal{M} (\vec{x})[\vec{y}]
=_{\bydefinition}
\left.
  \dd{}{s}
  \mathcal{M} (\vec{x} + s \vec{y})
\right|_{s=0}
$. If it is necessary to emphasize the variable against which we differentiate, we also write $\Diff[\vec{x}] \mathcal{M} (\vec{x})[\vec{y}]$ instead of $\Diff \mathcal{M} (\vec{x})[\vec{y}]$.
} 
of~$\mathcal{L}_{\Lambda}(\temp)$ reads
\begin{equation}
  \label{eq:48}
  \Diff \mathcal{L}_{\Lambda}(\temp) 
  \left[ 
    \vartheta 
  \right]
  =
  \int_{\Omega}
  \rho
  \cheatvol
  \left(
    \frac{1}{\temp}
    -
    \Lambda
  \right)
  \vartheta
  \,
  \cvolumee
  .
\end{equation}
The derivative vanishes in all possible directions $\vartheta$ provided that $\Lambda = \frac{1}{\temp}$. The Lagrange multiplier $\Lambda$ is a number, hence the temperature field $\temp$ at which the derivative vanishes must be a spatially homogeneous temperature field. Using the constraint $\nettenergy = \widehat{\nettenergy}$ we can therefore conclude that the temperature field $\temp$ that for all $\vartheta$ satisfies
\begin{equation}
  \label{eq:49}
  \Diff \mathcal{L}_{\Lambda}(\temp) 
  \left[ 
    \vartheta 
  \right]
  =
  0
\end{equation}
is the uniform temperature field $\temp_{\reference} = \tempbdr = \widehat{\temp}$. This calculation confirms the expected fact that the \emph{spatially homogeneous temperature field is the state that our thermodynamically isolated system wants to reach}.

\begin{figure}[h]
  \centering
  \subfloat[Plot of function $f(\temp)$ that appears as the integrand in~\eqref{eq:51}. \label{fig:auxiliary-a}]{\includegraphics[width=0.45\textwidth]{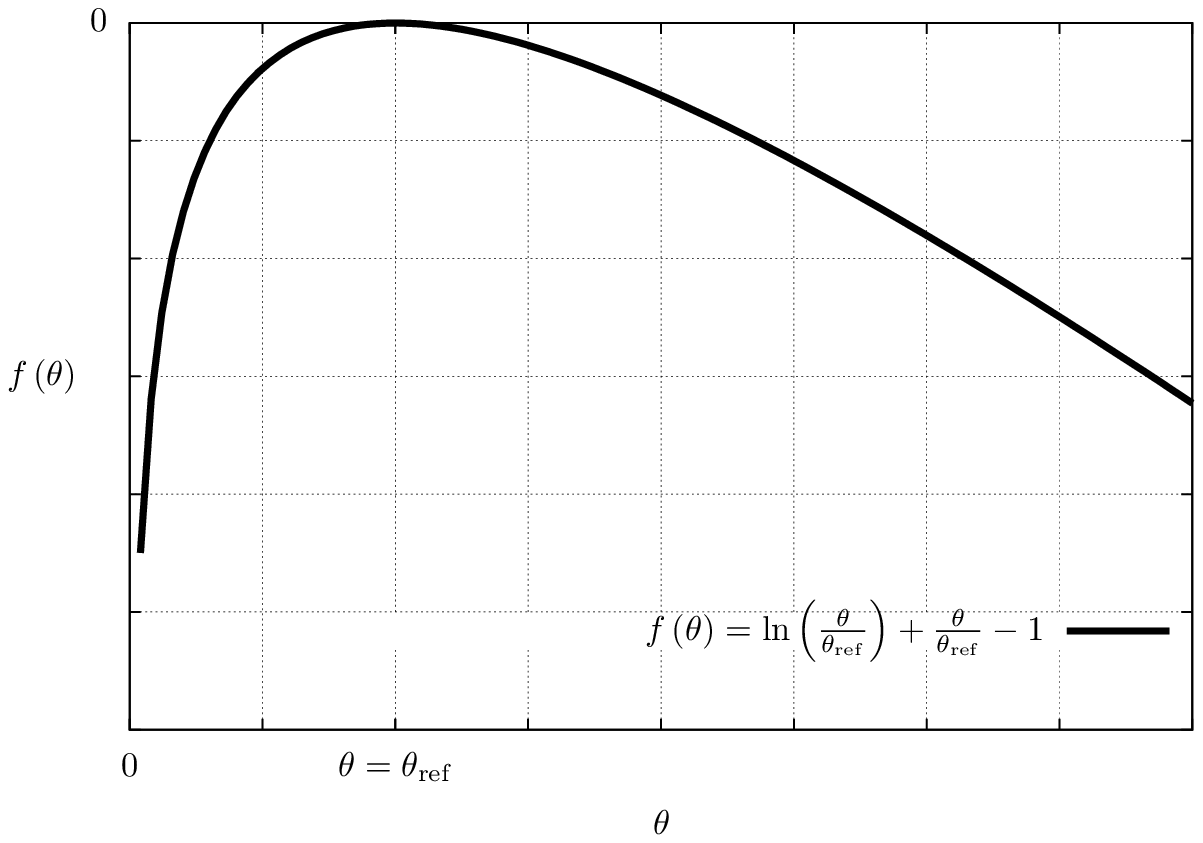}}
  \qquad
  \subfloat[Plot of function $g(\widetilde{\temp})$ that appears as the integrand in~\eqref{eq:94}. \label{fig:auxiliary-b}]{\includegraphics[width=0.45\textwidth]{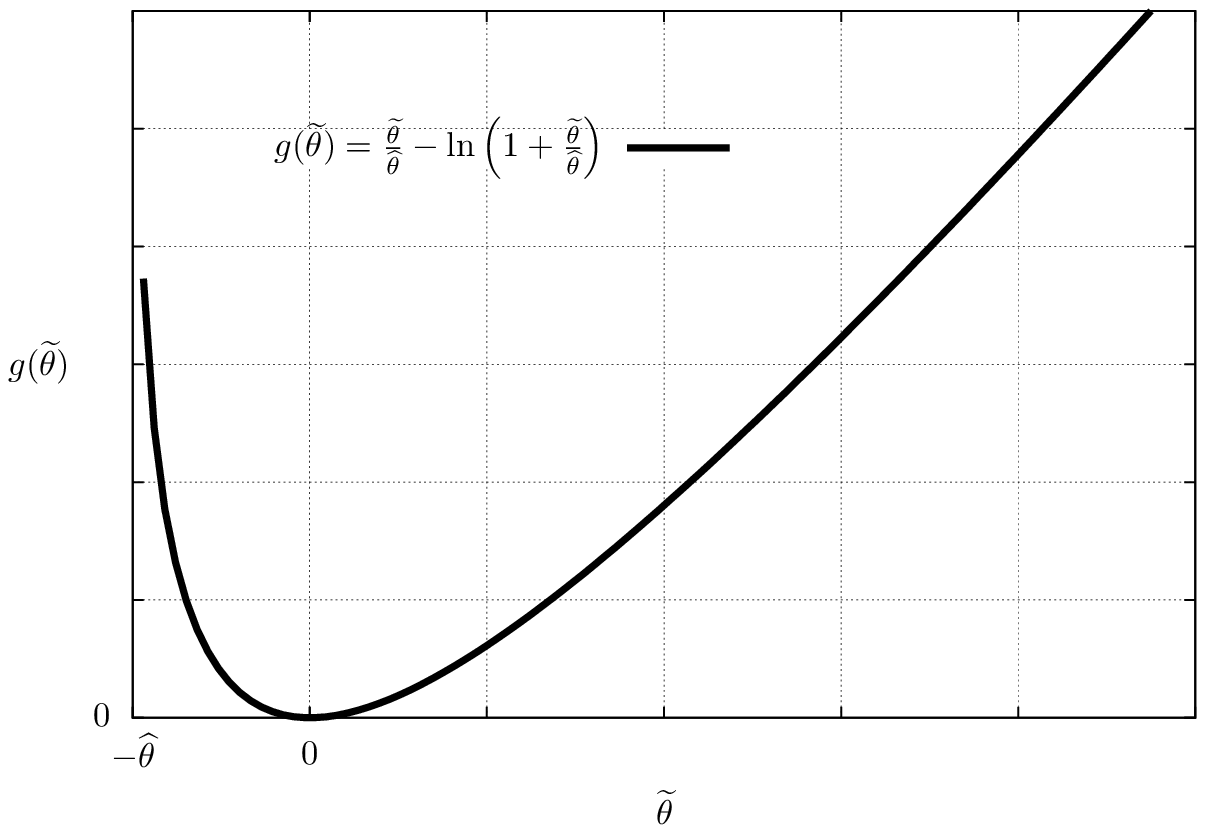}}
  \caption{Auxiliary functions.}
  \label{fig:auxiliary}
\end{figure}

Let us now exploit the fact that we know the value of the Lagrange multiplier $\Lambda$ in~\eqref{eq:47}, and let us investigate the functional
\begin{equation}
  \label{eq:50}
  \mathcal{L}_{\frac{1}{\widehat{\temp}}}(\temp) = _{\bydefinition} \netentropy - \frac{1}{\widehat{\temp}} \left( \nettenergy - \widehat{\nettenergy} \right).
\end{equation}
Explicit formula for the functional reads
\begin{equation}
  \label{eq:51}
  \mathcal{L}_{\frac{1}{\widehat{\temp}}}(\temp) 
  =
  \int_{\Omega}
  \rho
  \cheatvol 
  \left[
    \ln \left( \frac{\temp}{\widehat{\temp}} \right)
    -
    \frac{\temp}{\widehat{\temp}}
    +
    1
  \right]
  \,
  \cvolumee
  .
\end{equation}
(Recall that the reference temperature has been chosen as $\temp_{\reference} = \widehat{\temp}$.) We observe that function 
\begin{equation}
  \label{eq:52}
  f(\temp) 
  =_{\bydefinition}
  \ln \left( \frac{\temp}{\widehat{\temp}} \right)
  -
  \frac{\temp}{\widehat{\temp}}
  +
  1 
\end{equation}
is for $\temp > 0$ \emph{negative} whenever $\temp \not = \widehat{\temp}$, and it vanishes if and only if $\temp = \widehat{\temp}$. Further, this function is a concave function. The plot of the function $f$ is shown in Figure~\ref{fig:auxiliary-a}. 

Since the temperature field $\temp$ that solves~\eqref{eq:39} remains positive, we see that the functional $\mathcal{L}_{\frac{1}{\widehat{\temp}}}(\temp)$ is nonpositive for all possible solutions to~\eqref{eq:39}. Moreover, it vanishes if and only if $\temp = \widehat{\temp}$ everywhere in the domain $\Omega$. In other words it vanishes at the equilibrium rest state $\widehat{\temp}$, and it provides a \emph{control on the spatial variations of the temperature field} with respect to the equilibrium value. This means that the functional
\begin{equation}
  \label{eq:53}
  \mathcal{V}_{\mathrm{eq}} (\temp)=_{\bydefinition} -\mathcal{L}_{\frac{1}{\widehat{\temp}}}(\temp),
\end{equation}
that is
\begin{equation}
  \label{eq:54}
  \mathcal{V}_{\mathrm{eq}}(\temp)
  =
  -
  \int_{\Omega}
  \rho
  \cheatvol 
  \left[
    \ln \left( \frac{\temp}{\widehat{\temp}} \right)
    -
    \frac{\temp}{\widehat{\temp}}
    +
    1
  \right]
  \,
  \cvolumee
  ,
\end{equation}
is a good candidate for a Lyapunov functional characterising the stability of the equilibrium rest state $\widehat{\temp}$.

It remains to check that the time derivative of the proposed Lyapunov functional $\mathcal{V}_{\mathrm{eq}}$ is nonpositive provided that the temperature field evolves according to the governing equations~\eqref{eq:39}. First, we observe that in a thermodynamically isolated system we get
\begin{equation}
  \label{eq:55}
  \dd{\nettenergy}{t} = 0.
\end{equation}
This follows by the integration of~\eqref{eq:30} over the domain $\Omega$, and from the fact that energy flux $\efluxc$ vanishes on the boundary\footnote{In the case of a deformable body, one would also need to specify boundary conditions for the velocity/stress field that imply that the body is thermodynamically isolated. This is fulfilled provided that $\left. \vectordot{\left(\cstress \vec{v}\right)}{\vec{n}} \right|_{\partial \Omega} = 0$, which is granted for example, by the zero boundary velocity condition $\left. \vec{v} \right|_{\partial \Omega} = \vec{0}$.}
. Second, the entropy of the thermodynamically isolated system is a nondecreasing function, see~\eqref{eq:44}. Consequently, we see that
\begin{equation}
  \label{eq:56}
  \dd{\mathcal{V}_{\mathrm{eq}}}{t}
  =
  -
  \dd{}{t}
  \left[
    \netentropy - \frac{1}{\widehat{\temp}} \left( \nettenergy - \widehat{\nettenergy} \right)
  \right]
  =
  -
  \dd{\netentropy}{t}
  =
  -
  \int_{\Omega} \frac{\entprodctemp}{\temp} \, \cvolumee
  \leq
  0
  .
\end{equation}
Moreover, the derivative vanishes if and only if the given temperature field is spatially homogeneous. (Recall that $\entprodctemp= \kappa \frac{\absnorm{\nabla \temp}^2}{\temp}$, see~\eqref{eq:26}.) This concludes that~$\mathcal{V}_{\mathrm{eq}}$ is indeed a suitable Lyapunov functional characterising the stability of the equilibrium rest state, hence the equilibrium rest state is unconditionally asymptotically stable.

The fact that the functional of the type $\netentropy - \frac{1}{\tempbdr} \nettenergy$ can be used as a Lyapunov functional characterising the stability of the equilibrium rest state in a thermodynamically isolated system is well known, see~\cite{coleman.bd:on}, \cite{gurtin.me:thermodynamics}, \cite{silhavy.m:mechanics}, \cite{ericksen.jl:introduction} or \cite{grmela.m.ottinger.hc:dynamics}. In fact \cite{gurtin.me:thermodynamics} attributes this observation to~\cite{duhem.p:traite}, yet the core idea can be, for spatially homogeneous systems, found already in the works of~\cite{clausius.r:ueber} and \cite{gibbs.jw:equilibrium*1,gibbs.jw:equilibrium}. 

Interestingly, the functional $\netentropy - \frac{1}{\tempbdr} \nettenergy$ is not used or even mentioned in standard treatises on nonlinear stability analysis, see~\cite{joseph.dd:stability*1,joseph.dd:stability} or~\cite{straughan.b:energy}. This is in a sense natural, since these works are mostly focused on~\emph{thermodynamically open} systems, where the approach introduced in the seminal work by~\cite{coleman.bd:on} is largely inapplicable. On the other hand, this omission can be seen as an evidence of the perceived inapplicability of genuine thermodynamical concepts in the nonlinear stability analysis of \emph{thermodynamically open} systems. 

\subsubsection{Relation to the standard energy method}
\label{sec:relat-class-energy}
The constructed functional $\mathcal{V}_{\mathrm{eq}}$ coincides, up to a constant, with the standard functional $\mathcal{V}_{\mathrm{std}}$ introduced in Section~\ref{sec:stand-proof-uncond} provided that the temperature perturbation is small. Indeed, if $\frac{\temp}{\temp_{\reference}} \approx 1$, then $f(\temp) \approx \frac{\temp^2}{\temp_{\reference}^2}$, and consequently
\begin{equation}
  \label{eq:57}
  \mathcal{V}_{\mathrm{eq}}  \approx \mathcal{V}_{\mathrm{std}}. 
\end{equation}

Note also that the functional $\mathcal{V}_{\mathrm{eq}}$ can be seen, up to a constant factor, as the generalisation of the classical concept of \emph{exergy}/\emph{available energy}, see for example~\cite{bruges.ea:available}, to spatially inhomogeneous systems.

\subsection{Unconditional asymptotic stability of a general steady state in a thermodynamically open system}
\label{sec:general-steady-state}
Having identified a physically motivated Lyapunov functional for the stability analysis of the rest state in a thermodynamically isolated system, we can proceed with the stability analysis of steady solution in a \emph{thermodynamically open system}. 

\subsubsection{Governing equations for the non-equilibrium steady state}
\label{sec:govern-equat-non}
We consider the heat conduction problem in a rigid body with a given temperature value $\tempbdr$ on the boundary, where the temperature value $\tempbdr$ on the boundary can be \emph{position dependent}. (A part of the boundary can be kept at a different temperature than the other. A~good model problem is the heat conduction problem in a rod that has its ends kept at different temperatures.) This means that the analysis below \emph{is not restricted} to the setting of body ``immersed in a environment of [spatially uniform] temperature'', see~\cite{coleman.bd:on} 
and similar works such as \cite{gurtin.me:thermodynamics*1,gurtin.me:thermodynamics}. 

Let $\widehat{\temp}$ denote the steady solution to the boundary-value problem~\eqref{eq:6}, that is the temperature field $\widehat{\temp}$ solves the problem
\begin{subequations}
  \label{eq:non-equilibrium-rest-state}
  \begin{align}
    \label{eq:58}
    0 &= \divergence \left( \kappa \nabla \widehat{\temp} \right), \\
    \label{eq:59}
    \left. \widehat{\temp} \right|_{\partial \Omega} &= \tempbdr. 
  \end{align}
\end{subequations}
This is the steady solution whose stability we want to investigate.

Note that if the boundary condition $\tempbdr$ is \emph{spatially inhomogeneous}, then the solution $\widehat{\temp}$ is spatially inhomogeneous as well. Consequently, the entropy production~\eqref{eq:26} at the steady state $\widehat{\temp}$ is positive, which makes the widely used mathematical term \emph{equilibrium solution} for $\widehat{\temp}$ a bit problematic from the physical point of view. The system we are interested in is from the thermodynamical point of view \emph{out of thermodynamical equilibrium}. (Entropy is being produced.) Therefore $\widehat{\temp}$ is from this point of view a \emph{non-equilibrium steady state}.

\subsubsection{Governing equations for the perturbation}
\label{sec:govern-equat-pert-1}
The time evolution of the perturbed temperature field $\temp = \widehat{\temp} + \widetilde{\temp}$ is governed by equations~\eqref{eq:10}, that is
\begin{subequations}
  \label{eq:60}
  \begin{align}
    \label{eq:61}
    \rho \cheatvol \pd{\temp}{t} &= \divergence \left( \kappa \nabla \temp \right),  \\
    \label{eq:62}
    \left. \temp \right|_{\partial \Omega} &= \tempbdr, \\
    \label{eq:63}
    \left. \temp \right|_{t=0} &= \tempinit,
  \end{align}
\end{subequations}
where $\tempinit$ is an arbitrary initial condition. Consequently, the time evolution of the perturbation $\widetilde{\temp}$ is governed by
\begin{subequations}
  \label{eq:64}
  \begin{align}
    \label{eq:65}
    \rho \cheatvol \pd{\widetilde{\temp}}{t} &= \divergence \left( \kappa \nabla \widetilde{\temp} \right),  \\
    \label{eq:66}
    \left. \widetilde{\temp} \right|_{\partial \Omega} &= 0, \\
    \label{eq:67}
    \left. \widetilde{\temp} \right|_{t=0} &= \tempinit - \widehat{\temp}.
  \end{align}
\end{subequations}
The aim is to show that the perturbation $\widetilde{\temp}$ vanishes as time goes to infinity irrespective of the choice of the initial condition~$\tempinit$.

\subsubsection{Heuristics concerning the construction of a physically motivated Lyapunov functional}
\label{sec:heur-conc-constr}
Concerning the stability analysis of the steady state $\widehat{\temp}$ we again want to exploit the concept of Lyapunov functional. The following observation will be helpful. Let us assume that we have a quadratic positive definite functional defined on the real line, say 
\begin{equation}
  \label{eq:68}
  \mathcal{V}_{\mathrm{eq}}(\left. \widetilde{x}_{\mathrm{eq}} \right\| \widehat{x}_{\mathrm{eq}}) =_{\bydefinition} \widetilde{x}_{\mathrm{eq}}^2,
\end{equation}
that can be used as the Lyapunov functional for the stability analysis of an equilibrium rest state $\widehat{x}_{\mathrm{eq}}$. Here the perturbation with respect to the rest state is denoted as $\widetilde{x}_{\mathrm{eq}}$, and the complete perturbed field $x$ is defined as 
\begin{equation}
  \label{eq:69}
  x = \widehat{x}_{\mathrm{eq}} + \widetilde{x}_{\mathrm{eq}}. 
\end{equation}
Note that in terms of the complete perturbed field  $x$ we get $\mathcal{V}_{\mathrm{eq}}(\left. \widetilde{x}_{\mathrm{eq}} \right\| \widehat{x}_{\mathrm{eq}}) = (x - \widehat{x}_{\mathrm{eq}})^2$. Consequently, we also use, whenever appropriate, the notation  $\mathcal{V}_{\mathrm{eq}}(x) =_{\bydefinition} (x - \widehat{x}_{\mathrm{eq}})^2$ or 
\begin{equation}
  \label{eq:70}
  \mathcal{V}_{\mathrm{eq}}(x) =_{\bydefinition} \mathcal{V}_{\mathrm{eq}}(\left. \widetilde{x}_{\mathrm{eq}} \right\| \widehat{x}_{\mathrm{eq}}).
\end{equation}
The latter notation $\mathcal{V}_{\mathrm{eq}}(x)$ indicates that we are dealing with the complete perturbed field $x$, while the former notation~$\mathcal{V}_{\mathrm{eq}}(\left. \widetilde{x}_{\mathrm{eq}} \right\| \widehat{x}_{\mathrm{eq}})$ indicates that we are interested in the stability of the steady state $\widehat{x}_{\mathrm{eq}}$ subject to perturbations $\widetilde{x}_{\mathrm{eq}}$.

Now we want to construct a new functional $\mathcal{V}_{\mathrm{neq}} (\left. \widetilde{x}_{\mathrm{neq}} \right\| \widehat{x}_{\mathrm{neq}}) $ that could serve as a Lyapunov functional characterising the stability of the non-equilibrium steady state $\widehat{x}_{\mathrm{neq}}$. The point~$\widehat{x}_{\mathrm{neq}}$ represents the non-equilibrium steady state whose stability we are interested in, and $\widetilde{x}_{\mathrm{neq}}$ \emph{denotes the perturbation with respect to the nonequlibrium state}~$\widehat{x}_{\mathrm{neq}}$. The complete perturbed field $x$ is again composed of the perturbation $\widetilde{x}_{\mathrm{neq}}$ and the non-equilibrium steady state $\widehat{x}_{\mathrm{neq}}$,
\begin{equation}
  \label{eq:71}
  x = \widehat{x}_{\mathrm{neq}} + \widetilde{x}_{\mathrm{neq}}. 
\end{equation} We want the new functional $\mathcal{V}_{\mathrm{neq}}$ to vanish if the perturbation $\widetilde{x}_{\mathrm{neq}}$ vanishes, and to be positive otherwise. 

The new functional $\mathcal{V}_{\mathrm{neq}}$ can be constructed from $\mathcal{V}_{\mathrm{eq}}$ as follows. We ``subtract'' the tangent to the graph of the former functional $\mathcal{V}_{\mathrm{eq}}$ at the point~$\widehat{x}_{\mathrm{neq}}$ from the graph of the former functional $\mathcal{V}_{\mathrm{eq}}$. (See Figure~\ref{fig:construction} for a sketch of the construction.) In other words, the new functional $\mathcal{V}_{\mathrm{neq}}$ is defined as
\begin{subequations}
  \label{eq:72}
  \begin{equation}
    \label{eq:73}
    \mathcal{V}_{\mathrm{neq}}(\left. \widetilde{x}_{\mathrm{neq}} \right\| \widehat{x}_{\mathrm{neq}})
    =_{\bydefinition} 
    \mathcal{V}_{\mathrm{eq}}(\left. \widehat{x}_{\mathrm{neq}} + \widetilde{x}_{\mathrm{neq}} -  \widehat{x}_{\mathrm{eq}} \right\|  \widehat{x}_{\mathrm{eq}}) 
    - 
    \mathcal{V}_{\mathrm{eq}}(\left. \widehat{x}_{\mathrm{neq}} -  \widehat{x}_{\mathrm{eq}} \right\|  \widehat{x}_{\mathrm{eq}}) 
    - 
    \left. \dd{\mathcal{V}_{\mathrm{eq}}}{x} \right|_{x=\widehat{x}_{\mathrm{neq}}} \widetilde{x}_{\mathrm{neq}}
    ,
  \end{equation}
  or in other words as
  \begin{equation}
    \label{eq:74}
    \mathcal{V}_{\mathrm{neq}}(\left. \widetilde{x}_{\mathrm{neq}} \right\| \widehat{x}_{\mathrm{neq}})
    =_{\bydefinition} 
    \mathcal{V}_{\mathrm{eq}}(\widehat{x}_{\mathrm{neq}} + \widetilde{x}_{\mathrm{neq}}) 
    - 
    \mathcal{V}_{\mathrm{eq}}(\widehat{x}_{\mathrm{neq}}) 
    - 
    \left. \dd{\mathcal{V}_{\mathrm{eq}}}{x} \right|_{x=\widehat{x}_{\mathrm{neq}}} \widetilde{x}_{\mathrm{neq}}
    .
  \end{equation}
\end{subequations}
Formula~\eqref{eq:72} can be as well read as the ``remainder'' after subtracting the first order expansion of the original functional~$\mathcal{V}_{\mathrm{eq}}$ at the point $\widehat{x}_{\mathrm{neq}}$ from the functional $\mathcal{V}_{\mathrm{eq}}$.

In the case of functional~\eqref{eq:68} we get
\begin{equation}
  \label{eq:75}
    \mathcal{V}_{\mathrm{neq}}(\left. \widetilde{x}_{\mathrm{neq}} \right\| \widehat{x}_{\mathrm{neq}})
    =
    \left(\widehat{x}_{\mathrm{neq}} + \widetilde{x}_{\mathrm{neq}} - \widehat{x}_{\mathrm{eq}}\right)^2
    -
    \left(\widehat{x}_{\mathrm{neq}} - \widehat{x}_{\mathrm{eq}}\right)^2
    -
    2
    \left(
      \widehat{x}_{\mathrm{neq}} - \widehat{x}_{\mathrm{eq}}
    \right)
    \widetilde{x}_{\mathrm{neq}}
    =
    \widetilde{x}_{\mathrm{neq}}^2
    .
\end{equation}
Apparently, the newly constructed functional $\mathcal{V}_{\mathrm{neq}}(\left. \widetilde{x}_{\mathrm{neq}} \right\| \widehat{x}_{\mathrm{neq}})$ is positive provided $\widetilde{x}_{\mathrm{neq}} \not = 0$. Moreover, it vanishes at $\widetilde{x}_{\mathrm{neq}} = 0$, that is if the state of the system $x = \widehat{x}_{\mathrm{neq}} + \widetilde{x}_{\mathrm{neq}}$ is identical to the chosen non-equilibrium steady state~$\widehat{x}_{\mathrm{neq}}$. Consequently, $\mathcal{V}_{\mathrm{neq}}(\left. \widetilde{x}_{\mathrm{neq}} \right\| \widehat{x}_{\mathrm{neq}})$ is a reasonable \emph{guess} concerning the Lyapunov functional suitable for the stability analysis of the non-equilibrium state~$\widehat{x}_{\mathrm{neq}}$. 

It remains to show that the newly constructed functional is decreasing along the trajectories predicted by the corresponding governing equations for $x$. If this can be shown, then the newly constructed functional is indeed a Lyapunov functional suitable for the analysis of the stability of the non-equilibrium state $\widehat{x}_{\mathrm{neq}}$. In this heuristic argument we however do not consider any underlying dynamical system, hence we can not proceed further in the study of the property $\dd{\mathcal{V}_{\mathrm{neq}}(\left. \widetilde{x}_{\mathrm{neq}} \right\| \widehat{x}_{\mathrm{neq}})}{t} \leq{0}$.

\begin{figure}[h]
  \centering
  \includegraphics[width=0.5\textwidth]{figures/sketch/lyapunov-nonequilibrium}
  \caption{Construction of the Lyapunov functional $\mathcal{V}_{\mathrm{neq}}(\left. \widetilde{x}_{\mathrm{neq}} \right\| \widehat{x}_{\mathrm{neq}})$ for a non-equilibrium state $\widehat{x}_{\mathrm{neq}}$ from the Lyapunov functional $\mathcal{V}_{\mathrm{eq}}(\left. \widetilde{x}_{\mathrm{eq}} \right\| \widehat{x}_{\mathrm{eq}})$ for the rest state $\widehat{x}_{\mathrm{eq}}$.}
  \label{fig:construction}
\end{figure}

We note that the outlined construction is quite general, and it can be easily extended to the multidimensional or even infinite-dimensional setting. The key property that guarantees a meaningful outcome of the outlined construction of $\mathcal{V}_{\mathrm{neq}}$ is the convexity of the functional characterising the stability of the equilibrium rest state $\mathcal{V}_{\mathrm{eq}}$. The origins of the outlined construction can be, to our best knowledge, traced back to~\cite{ericksen.jl:thermo-kinetic}.

\subsubsection{Construction of a physically motivated Lyapunov functional -- general remarks}
\label{sec:constr-lyap-funct}
Let us now follow the outlined heuristic in the case of dynamical systems in continuum thermodynamics, and especially in the case of heat conduction. The Lyapunov functional $\mathcal{V}_{\mathrm{eq}}$ for the \emph{equilibrium rest state} in a thermodynamically closed system has been identified in~\eqref{eq:53}, 
\begin{equation}
  \label{eq:76}
  \mathcal{V}_{\mathrm{eq}} 
  =
  -
  \left\{
    \netentropy - \frac{1}{\widehat{\temp}} \left( \nettenergy - \widehat{\nettenergy} \right)
  \right\}
  .
\end{equation}
However, if we want to use~\eqref{eq:76} as a building block for a Lyapunov functional characterising the stability of a steady non-equilibrium state with a spatially inhomogeneous temperature field $\widehat{\temp}$, we see that~\eqref{eq:76} does not define a functional. It does not assign a \emph{real number} to the given state of the system (temperature field). (While $\netentropy$ and $\nettenergy$ in~\eqref{eq:76} are numbers even if one deals with spatially inhomogeneous temperature field, the factor $\frac{1}{\widehat{\temp}}$ is in the spatially inhomogenous setting a position dependent function.) This can be fixed if we realise that the Lyapunov functional characterising the stability of the \emph{equilibrium rest state} can be rewritten as
\begin{equation}
  \label{eq:77}
  \mathcal{V}_{\mathrm{eq}} 
  =
  -
  \frac{1}{\widehat{\temp}}
  \left\{
    \widehat{\temp}
    \netentropy
    -
    \left( \nettenergy - \widehat{\nettenergy} \right)
  \right\}
  =
  -
  \frac{1}{\widehat{\temp}}
  \left\{
    \int_{\Omega}
    \widehat{\temp}
    \rho
    \entropy
    \,
    \cvolumee
    -
    \left(
      \int_{\Omega}
      \rho
      \left[
        \frac{1}{2} \absnorm{\vec{v}}^2
        +
        \ienergy
      \right]
      \,
      \cvolumee
      -
      \widehat{\nettenergy}
    \right)
  \right\}
  ,
\end{equation}
where we have used the definition of the net total energy and the net entropy, see~\eqref{eq:net-total-energy-and-entropy}. The factor $\frac{1}{\widehat{\temp}}$ in~\eqref{eq:77} is immaterial in the stability analysis of a spatially homogeneous equilibrium rest state $\widehat{\temp}$. Indeed the modified Lyapunov functional   
\begin{equation}
  \label{eq:78}
  \mathcal{V}_{\mathrm{meq}} 
  =_{\bydefinition}
  -
  \left\{
    \int_{\Omega}
    \widehat{\temp}
    \rho
    \entropy
    \,
    \cvolumee
    -
    \left(
      \int_{\Omega}
      \rho
      \left[
        \frac{1}{2} \absnorm{\vec{v}}^2
        +
        \ienergy
      \right]
      \,
      \cvolumee
      -
      \widehat{\nettenergy}
    \right)
  \right\}
\end{equation}
can serve as well as the original Lyapunov functional~\eqref{eq:76} in the stability analysis of the spatially homogeneous equilibrium rest state. 

Introducing the notation
\begin{equation}
  \label{eq:79}
  \netentropy_{\widehat{\temp}} 
  =
  _{\bydefinition}
  \int_{\Omega}
  \rho
  \widehat{\temp}
  \entropy
  \,
  \cvolumee
  ,
\end{equation}
we see that~\eqref{eq:78} can be rewritten as
\begin{equation}
  \label{eq:80}
  \mathcal{V}_{\mathrm{meq}} 
  =
  -
  \left\{
    \netentropy_{\widehat{\temp}} 
    -
    \left( \nettenergy - \widehat{\nettenergy} \right)
  \right\}
  .
\end{equation}
Note that the definition~\eqref{eq:80} of $\mathcal{V}_{\mathrm{meq}}$ is general enough to be applicable whenever one deals with a continuous medium with a well defined specific Helmholtz free energy. It is by no means restricted to the specific problem of heat conduction in a rigid body. 

The benefit of using $\mathcal{V}_{\mathrm{meq}}$ instead of $\mathcal{V}_{\mathrm{eq}}$ lies in the fact that the temperature field $\widehat{\temp}$ is now placed under the integration sign, hence $\mathcal{V}_{\mathrm{meq}}$ defines a functional even if $\widehat{\temp}$ is a spatially inhomogeneous temperature field. This subtlety does not matter if $\widehat{\temp}$ is a constant temperature field. On the other hand, it allows us to use the Lyapunov functional $\mathcal{V}_{\mathrm{meq}}$ as functional that serves as a building block in the construction of the Lyapunov functional $\mathcal{V}_{\mathrm{neq}}$ for the non-equlibrium steady state~$\widehat{\temp}$.

\subsubsection{Construction of a physically motivated Lyapunov functional -- heat conduction in a rigid body}
\label{sec:constr-phys-motiv-1}
Now we are in a position to follow the construction outlined in Section~\ref{sec:heur-conc-constr} and Section~\ref{sec:constr-lyap-funct} for. Let $\vec{W}$ denote the vector of state variables, and let~$\widetilde{\vec{W}}$ and~$\widehat{\vec{W}}$ denote a perturbation and a non-equilibrium steady state respectively. The candidate for Lyapunov functional is defined as
\begin{subequations}
  \label{eq:82}
\begin{equation}
  \label{eq:81}
  \mathcal{V}_{\mathrm{neq}}
  \left(
    \left.
      \widetilde{\vec{W}}
    \right\|
    \widehat{\vec{W}}
  \right)
  =
  -
  \left\{
    \entropyrellp_{\widehat{\temp}}(\left. \widetilde{\vec{W}} \right\| \widehat{\vec{W}})
    -
    \energyrellp  (\left. \widetilde{\vec{W}} \right\| \widehat{\vec{W}})
  \right\}
  ,
\end{equation}
where
  \begin{align}
    \label{eq:83}
    \entropyrellp_{\widehat{\temp}}(\left. \widetilde{\vec{W}} \right\| \widehat{\vec{W}})
    &=
    _{\bydefinition}
    \netentropy_{\widehat{\temp}} 
    \left(
      \widehat{\vec{W}}
      +
      \widetilde{\vec{W}}
    \right)
    -
    \netentropy_{\widehat{\temp}} 
    \left(
      \widehat{\vec{W}}
    \right)
    -
    \left.
      \Diff[\vec{W}] \netentropy_{\widehat{\temp}}
      \left(
        \vec{W}
      \right)
    \right|_{\vec{W}  = \widehat{\vec{W}}}
    \left[
      \widetilde{\vec{W}}
    \right]
    ,
    \\
    \label{eq:84}
    \energyrellp  (\left. \widetilde{\vec{W}} \right\| \widehat{\vec{W}})
    &=
    _{\bydefinition}
    \nettenergy 
    \left(
      \widehat{\vec{W}}
      +
      \widetilde{\vec{W}}
    \right)
    -
    \nettenergy 
    \left(
      \widehat{\vec{W}}
    \right)
    -
    \left.
      \Diff[\vec{W}] \nettenergy
      \left(
        \vec{W}
      \right)
    \right|_{\vec{W}  = \widehat{\vec{W}}}
    \left[
      \widetilde{\vec{W}}
    \right]
    ,
  \end{align}
\end{subequations}
and the functionals~$\netentropy_{\widehat{\temp}} \left(\vec{W}\right)$ and $\nettenergy\left(\vec{W}\right)$ are defined as
\begin{subequations}
  \label{eq:85}
  \begin{align}
    \label{eq:86}
    \netentropy_{\widehat{\temp}} \left(\vec{W}\right)
    &=
    _{\bydefinition}
    \int_{\Omega}
    \rho
    \widehat{\temp}
    \entropy(\vec{W})
    \,
    \cvolumee
    ,
    \\
    \label{eq:87}
    \nettenergy\left(\vec{W}\right)
    &=
    _{\bydefinition}
    \int_{\Omega}
    \rho
    \ienergy(\vec{W})
    \,
    \cvolumee
    .
  \end{align}
\end{subequations}
In~\eqref{eq:81} we have rewritten~\eqref{eq:72} in the infinite-dimensional setting, meaning that the derivative has been replaced by the G\^ateaux derivative, see Footnote~\ref{fn:1} for the definition.

In our case the specific entropy $\entropy$ and the specific internal energy $\ienergy$ are given by the formulae
\begin{equation}
  \label{eq:88}
  \entropy(\vec{W})
  =
  \cheatvol \ln \left( \frac{\temp}{\temp_{\reference}} \right) 
  ,
  \qquad
  \ienergy(\vec{W})
  =
  \cheatvol
  \temp
  ,
\end{equation}
see~\eqref{eq:24} and \eqref{eq:25}, and the only state variable is the temperature field, that is $\vec{W} =_{\bydefinition} \temp$, $\widehat{\vec{W}} =_{\bydefinition} \widehat{\temp}$ and $\widetilde{\vec{W}} =_{\bydefinition} \widetilde{\temp}$. The G\^ateaux derivatives of the functionals~$\netentropy_{\widehat{\temp}}(\vec{W})$ and $\nettenergy (\vec{W})$ read\footnote{Note that the differentiation in~\eqref{eq:90} requires one to vary only $\vec{W}$. The factor $\widehat{\temp}$ in $ \netentropy_{\widehat{\temp}} (\vec{W})$ is left intact although we are differentiating with respect to the temperature. We get
$
\left.
  \Diff[\vec{W}] \netentropy_{\widehat{\temp}} (\vec{W})
\right|_{
  \vec{W} = \widehat{\vec{W}}
}
\left[
  \widetilde{\vec{W}}
\right]
=
\left.
  \left\{
    \dd{}{s}
    \int_{\Omega}
    \rho
    \widehat{\temp}
    \cheatvol \ln \left( \frac{\widehat{\temp} + s \widetilde{\temp}}{\temp_{\reference}} \right) 
    \,
    \cvolumee
  \right\}
\right|_{s=0}
=
\int_{\Omega}
\rho
\cheatvol \widetilde{\temp}
\,
\cvolumee
$.
}
\begin{subequations}
  \label{eq:89}
  \begin{align}
    \label{eq:90}
    \left.
      \Diff[\vec{W}] \netentropy_{\widehat{\temp}} (\vec{W})
    \right|_{
      \vec{W} = \widehat{\vec{W}}
    }
    \left[
      \widetilde{\vec{W}}
    \right]
    &=
    \int_{\Omega}
    \rho
    \cheatvol \widetilde{\temp}
    \,
    \cvolumee
    ,
    \\
    \label{eq:91}
    \left.
      \Diff[\vec{W}] \nettenergy (\vec{W})
    \right|_{
      \vec{W} = \widehat{\vec{W}}
    }
    \left[
      \widetilde{\vec{W}}
    \right]
    &=
    \int_{\Omega}
    \rho
    \cheatvol \widetilde{\temp}
    \,
    \cvolumee
    .
  \end{align}
\end{subequations}
This calculation reveals that 
\begin{subequations}
  \label{eq:92}
  \begin{align}
    \label{eq:93}
    \energyrellp  (\left. \widetilde{\vec{W}} \right\| \widehat{\vec{W}}) 
    &= 
    0
    ,
    \\
    \label{eq:94}
    -
    \entropyrellp_{\widehat{\temp}}(\left. \widetilde{\vec{W}} \right\| \widehat{\vec{W}})
    &=
    \int_{\Omega}
    \rho
    \cheatvol
    \widehat{\temp}
    \left[
      \frac{\widetilde{\temp}}{\widehat{\temp}}
      -
      \ln \left( 1 + \frac{\widetilde{\temp}}{\widehat{\temp}} \right) 
    \right]
    \,
    \cvolumee
    .
  \end{align}
\end{subequations}

The function in the integrand in~\eqref{eq:94}, 
\begin{equation}
  \label{eq:95}
  g(\widetilde{\temp}) 
  =_{\bydefinition}
  \left[
    \frac{\widetilde{\temp}}{\widehat{\temp}}
    -
    \ln \left( 1 + \frac{\widetilde{\temp}}{\widehat{\temp}} \right) 
  \right]
  ,
\end{equation}
see the plot shown in Figure~\ref{fig:auxiliary-b}, is well defined for $\widetilde{\temp} > -\widehat{\temp}$. The governing equation~\eqref{eq:61} guarantees that the temperature field $\temp = \widetilde{\temp} + \widehat{\temp}$ remains positive provided that the initial temperature field $\tempinit$ is positive, see for example~\cite{friedman.a:partial}, \cite{ladyzhenskaya.oa.solonnikov.va.ea:linear} or \cite{lieberman.gm:second}. The pointwise values of the temperature perturbation~$\widetilde{\temp}$ therefore remain in the interval $\left( -\widehat{\temp}, +\infty \right)$, and the integrand remains well defined for any temperature field predicted by the corresponding governing equation. Moreover, the function $g$ and hence the integrand in \eqref{eq:94} is positive whenever $\widetilde{\temp} \not=0$, and it vanishes at~$\widetilde{\temp} = 0$. 

This implies that the functional $\mathcal{V}_{\mathrm{neq}}$ given by the explicit formula
\begin{equation}
  \label{eq:96}
  \mathcal{V}_{\mathrm{neq}}
  \left(
    \left.
      \widetilde{\vec{W}}
    \right\|
    \widehat{\vec{W}}
  \right)
  =
  \int_{\Omega}
  \rho
  \cheatvol
  \widehat{\temp}
  \left[
    \frac{\widetilde{\temp}}{\widehat{\temp}}
    -
    \ln \left( 1 + \frac{\widetilde{\temp}}{\widehat{\temp}} \right) 
  \right]
  \,
  \cvolumee
\end{equation}
is well defined and nonnegative for any achievable temperature field $\widehat{\temp} + \widetilde{\temp}$, and it vanishes if and only if $\widetilde{\temp}=0$ in~$\Omega$. Therefore, we can conclude that the functional~$\mathcal{V}_{\mathrm{neq}}$ is a good candidate for a Lyapunov functional characterising the stability of non-equilibrium steady state $\widehat{\temp}$. In particular, it provides a control on the \emph{spatial inhomogeneity} of the perturbation $\widetilde{\temp}$. If~$\mathcal{V}_{\mathrm{neq}}$ is equal to zero, then the perturbation $\widetilde{\temp}$ vanishes everywhere in the whole domain $\Omega$. 

\subsubsection{Time derivative of the Lyapunov functional}
\label{sec:time-deriv-prop}
We have seen that the proposed functional $\mathcal{V}_{\mathrm{neq}}$ is a suitable measure of the size of the perturbation $\widetilde{\temp}$. It remains to show that the time derivative is nonpositive,$\dd{\mathcal{V}_{\mathrm{neq}}}{t} \leq 0$. The time derivative must be evaluated with the help of the governing equations~\eqref{eq:64} for the perturbation $\widetilde{\temp}$. The key difficulty in evaluating the time derivative is that the heat flux \emph{does not vanish on the boundary} since we are dealing with a thermodynamically \emph{open} system. In particular, we can not exploit the equalities $\left. \vectordot{\nabla \widehat{\temp}}{\vec{n}} \right|_{\partial \Omega} = 0$ or $\left. \vectordot{\nabla \widetilde{\temp}}{\vec{n}} \right|_{\partial \Omega} = 0$ as in the case of a thermodynamically \emph{isolated} system. In general we must be able to handle the situation where
\begin{equation}
  \label{eq:97}
  \left. \vectordot{\nabla \widetilde{\temp}}{\vec{n}} \right|_{\partial \Omega} \not = 0.
\end{equation}

Since the Lyapunov functional $\mathcal{V}_{\mathrm{neq}}$ includes the term~\eqref{eq:79} that is defined in terms of the specific entropy, it turns out that it is convenient to formulate governing equations also for the relative specific entropy $\widetilde{\entropy} =_{\bydefinition} \entropy - \widehat{\entropy}$, that is
\begin{equation}
  \label{eq:98}
  \widetilde{\entropy} =_{\bydefinition} \entropy(\widehat{\vec{W}} + \widetilde{\vec{W}}) - \entropy(\widehat{\vec{W}}).
\end{equation}
This quantity measures the difference between the specific entropy at the perturbed state $\entropy(\widehat{\vec{W}} + \widetilde{\vec{W}})$ and the specific entropy at the non-equilibrium steady state $\entropy(\widehat{\vec{W}})$. In our case the explicit formula for $\widetilde{\entropy}$ in terms of temperature reads
\begin{equation}
  \label{eq:99}
  \widetilde{\entropy} = \cheatvol \ln \left( 1 + \frac{\widetilde{\temp}}{\widehat{\temp}}\right),
\end{equation}
and the evolution equation for $\widetilde{\entropy}$ is 
\begin{equation}
  \label{eq:100}
  \rho \pd{\widetilde{\entropy}}{t}
  =
  \frac{\kappa}{\cheatvol^2}
  \vectordot{\nabla \widetilde{\entropy}}{\nabla \widetilde{\entropy}}
  +
  \frac{2 \kappa}{\cheatvol^2}
  \vectordot{\nabla \widetilde{\entropy}}{\nabla \widehat{\entropy}}
  +
  \divergence
  \left(
    \frac{\kappa}{\cheatvol}
    \nabla \widetilde{\entropy}
  \right)
  .
\end{equation}
Equation~\eqref{eq:100} follows via subtracting the evolution equation for the specific entropy~\eqref{eq:31} formulated for the perturbed entropy field $\entropy = \widehat{\entropy} + \widetilde{\entropy} = \entropy(\widehat{\vec{W}} + \widetilde{\vec{W}})$ 
\begin{equation}
  \label{eq:101}
  \rho
  \pd{}{t}
  \left(
    \widehat{\entropy} + \widetilde{\entropy}
  \right)
  =
  \kappa \frac{\absnorm{\nabla \left(\widehat{\temp} + \widetilde{\temp} \right)}^2}{ \left( \widehat{\temp} + \widetilde{\temp} \right) ^2}
  +
  \divergence
  \left\{
    \kappa
    \frac{\nabla \left( \widehat{\temp} + \widetilde{\temp} \right)}{\widehat{\temp} + \widetilde{\temp}}
  \right\}
\end{equation}
from the equation~\eqref{eq:31} formulated for the non-equilibrium steady state $\entropy = \widehat{\entropy} = \entropy(\widehat{\vec{W}})$,
\begin{equation}
  \label{eq:102}
  \rho
  \pd{\widehat{\entropy}}{t}
  =
  \kappa \frac{\absnorm{\nabla \widehat{\temp}}^2}{\widehat{\temp}^2}
  +
  \divergence
  \left\{
    \kappa
    \frac{\nabla \widehat{\temp}}{\widehat{\temp}}
  \right\}
  .
\end{equation}
We also note that $\widetilde{\entropy} = 0$ whenever $\widetilde{\temp}=0$, hence 
\begin{equation}
  \label{eq:103}
  \left. \widetilde{\entropy} \right|_{\partial \Omega} = 0.
\end{equation}

In calculating the time derivative $\dd{\mathcal{V}_{\mathrm{neq}}}{t}$, we can either directly differentiate formula~\eqref{eq:96}, or we can proceed as follows
\begin{multline}
  \label{eq:104}
  \dd{}{t}
  \mathcal{V}_{\mathrm{neq}}
  \left(
    \left.
      \widetilde{\vec{W}}
    \right\|
    \widehat{\vec{W}}
  \right)
  =
  -
  \dd{}{t}
  \entropyrellp_{\widehat{\temp}}(\left. \widetilde{\vec{W}} \right\| \widehat{\vec{W}})
  +
  \dd{}{t}
  \nettenergy 
  \left(
    \widehat{\vec{W}}
    +
    \widetilde{\vec{W}}
  \right)
  -
  \dd{}{t}
  \nettenergy 
  \left(
    \widehat{\vec{W}}
  \right)
  -
  \dd{}{t}
  \int_{\Omega}
  \rho
  \cheatvol \widetilde{\temp}
  \,
  \cvolumee
  \\
  =
  -
  \dd{}{t}
  \int_{\Omega}
  \rho
  \widehat{\temp}
  \widetilde{\entropy}
  \,
  \cvolumee
  +
  \dd{}{t}
  \int_{\Omega}
  \rho
  \cheatvol \widetilde{\temp}
  \,
  \cvolumee 
  +
  \dd{}{t}
  \nettenergy 
  \left(
    \widehat{\vec{W}}
    +
    \widetilde{\vec{W}}
  \right)
  -
  \dd{}{t}
  \nettenergy 
  \left(
    \widehat{\vec{W}}
  \right)
  -
  \dd{}{t}
  \int_{\Omega}
  \rho
  \cheatvol \widetilde{\temp}
  \,
  \cvolumee
  \\
  =
  -
  \dd{}{t}
  \int_{\Omega}
  \rho
  \widehat{\temp}
  \widetilde{\entropy}
  \,
  \cvolumee
  +
  \dd{}{t}
  \nettenergy 
  \left(
    \widehat{\vec{W}}
    +
    \widetilde{\vec{W}}
  \right)
  -
  \dd{}{t}
  \nettenergy 
  \left(
    \widehat{\vec{W}}
  \right)
  .
\end{multline}
The time derivatives of the net total energy $\nettenergy$ can be evaluated using the governing equation for the energy. We get
\begin{subequations}
  \label{eq:105}
  \begin{align}
    \label{eq:106}
    \dd{}{t}
    \nettenergy 
    \left(
      \widehat{\vec{W}}
      +
      \widetilde{\vec{W}}
    \right)
    &=
    \int_{\Omega}
    \divergence \left\{ \kappa \nabla \left(\widehat{\temp} + \widetilde{\temp} \right) \right\}
    \,
    \cvolumee
    ,
    \\
    \label{eq:107}
    \dd{}{t}
    \nettenergy 
    \left(
      \widehat{\vec{W}}
    \right)
    &=
    \int_{\Omega}
    \divergence \left\{ \kappa \nabla \widehat{\temp} \right\}
    \,
    \cvolumee
    ,
  \end{align}
\end{subequations}
which is a straightforward consequence of~\eqref{eq:30} and the integration over the domain $\Omega$. Using~\eqref{eq:105} in~\eqref{eq:104} yields
\begin{equation}
  \label{eq:108}
  \dd{}{t}
  \mathcal{V}_{\mathrm{neq}}
  \left(
    \left.
      \widetilde{\vec{W}}
    \right\|
    \widehat{\vec{W}}
  \right)
  =
  -
  \dd{}{t}
  \int_{\Omega}
  \rho
  \widehat{\temp}
  \widetilde{\entropy}
  \,
  \cvolumee
  +
  \int_{\Omega}
  \divergence \left( \kappa \nabla \widetilde{\temp} \right)
  \,
  \cvolumee
  =
  -
  \int_{\Omega}
  \rho
  \widehat{\temp}
  \pd{\widetilde{\entropy}}{t}
  \,
  \cvolumee
  +
  \int_{\Omega}
  \divergence \left( \kappa \nabla \widetilde{\temp} \right)
  \,
  \cvolumee
  .
\end{equation}
(Let us again recall that $\widehat{\temp}$ is the non-equilibrium \emph{steady} state, that is $\pd{\widehat{\temp}}{t}=0$.) Next we use the evolution equation for $\widetilde{\entropy}$, see~\eqref{eq:100}, and we substitute into the first term in~\eqref{eq:108}. We get
\begin{equation}
  \label{eq:109}
  \dd{}{t}
  \mathcal{V}_{\mathrm{neq}}
  \left(
    \left.
      \widetilde{\vec{W}}
    \right\|
    \widehat{\vec{W}}
  \right)
  =
  -
  \int_{\Omega}
  \frac{\kappa}{\cheatvol^2}
  \widehat{\temp} \vectordot{\nabla \widetilde{\entropy}}{\nabla \widetilde{\entropy}}
  \,
  \cvolumee
  -
  \int_{\Omega}
  \frac{2 \kappa}{\cheatvol^2}
  \widehat{\temp} \vectordot{\nabla \widetilde{\entropy}}{\nabla \widehat{\entropy}}
  \,
  \cvolumee
  -
  \int_{\Omega}
  \widehat{\temp}
  \divergence
  \left(
    \frac{\kappa}{\cheatvol}
    \nabla \widetilde{\entropy}
  \right)
  \,
  \cvolumee
  +
  \int_{\Omega}
  \divergence \left( \kappa \nabla \widetilde{\temp} \right)
  \,
  \cvolumee
  .
\end{equation}

Apparently, the first term in~\eqref{eq:109} is in the leading order quadratic in the perturbation and it is nonpositive. The aim is to manipulate the remaining terms in such a way that the complete right-hand side of~\eqref{eq:109} is also in the leading order quadratic in the perturbation. This must be possible, since the Lyapunov functional
$  
\mathcal{V}_{\mathrm{neq}}
\left(
  \left.
    \widetilde{\vec{W}}
  \right\|
  \widehat{\vec{W}}
\right)
$
that is being differentiated with respect to time is in the leading order quadratic in the perturbation, and the governing equation for $\widetilde{\entropy}$ or $\widetilde{\temp}$ respectively does not contain a zeroth order term. 

In our case we can in fact show that the last three terms on the right-hand side of~\eqref{eq:109} vanish, and that the only term that remains on the right-hand side of~\eqref{eq:109} is negative for all nonzero perturbations, hence we get an unconditional stability result. Dealing with a counterpart of~\eqref{eq:109} in a more complex setting than the heat conduction problem one can of course expect the presence of terms that do not have a definite sign. Consequently, the right-hand side of~\eqref{eq:109} will be nonpositive only if the additional terms can be bounded by the nonpositive terms. This will lead to conditional stability results. 

Let us first investigate the last two terms in~\eqref{eq:109}. We get
\begin{equation}
  \label{eq:110}
  -
  \int_{\Omega}
  \widehat{\temp}
  \divergence
  \left(
    \frac{\kappa}{\cheatvol}
    \nabla \widetilde{\entropy}
  \right)
  \,
  \cvolumee
  +
  \int_{\Omega}
  \divergence \left( \kappa \nabla \widetilde{\temp} \right)
  \,
  \cvolumee
  =
  -
  \int_{\Omega}
  \divergence
  \left(
    \widehat{\temp}
    \frac{\kappa}{\cheatvol}
    \nabla \widetilde{\entropy}
  \right)
  \,
  \cvolumee
  +
  \int_{\Omega}
  \frac{\kappa}{\cheatvol}
  \vectordot{\nabla \widehat{\temp}}{\nabla \widetilde{\entropy}}
  \,
  \cvolumee
  +
  \int_{\Omega}
  \divergence \left( \kappa \nabla \widetilde{\temp} \right)
  \,
  \cvolumee
  .
\end{equation}
The definition of $\widetilde{\entropy}$, see~\eqref{eq:99}, implies the following formula for the gradient of $\widetilde{\entropy}$,
\begin{equation}
  \label{eq:111}
  \nabla \widetilde{\entropy}
  =
  \cheatvol \frac{1}{1 + \frac{\widetilde{\temp}}{\widehat{\temp}}}
  \nabla 
  \left( 
    \frac{\widetilde{\temp}}{\widehat{\temp}}
  \right)
  =
  \cheatvol 
  \frac{1}{\widehat{\temp} + \widetilde{\temp}} \nabla \widetilde{\temp}
  -
  \cheatvol 
  \frac{\widetilde{\temp}}{\widehat{\temp} \left( \widehat{\temp} + \widetilde{\temp} \right)} \nabla \widehat{\temp}.
\end{equation}
Using~\eqref{eq:111} in the first term on the right-hand side of~\eqref{eq:110} yields
\begin{multline}
  \label{eq:112}
  -
  \int_{\Omega}
  \divergence
  \left(
    \widehat{\temp}
    \frac{\kappa}{\cheatvol}
    \nabla \widetilde{\entropy}
  \right)
  \,
  \cvolumee
  +
  \int_{\Omega}
  \frac{\kappa}{\cheatvol}
  \vectordot{\nabla \widehat{\temp}}{\nabla \widetilde{\entropy}}
  \,
  \cvolumee
  +
  \int_{\Omega}
  \divergence \left( \kappa \nabla \widetilde{\temp} \right)
  \,
  \cvolumee
  =
  \int_{\Omega}
  \frac{\kappa}{\cheatvol}
  \vectordot{\nabla \widehat{\temp}}{\nabla \widetilde{\entropy}}
  \,
  \cvolumee
  +
  \int_{\Omega}
  \divergence
  \left\{
     \kappa 
     \left(
       1
       -
       \frac{\widehat{\temp}}{\widehat{\temp} + \widetilde{\temp}}
     \right)
     \nabla \widetilde{\temp}
     +
     \kappa 
     \frac{\widetilde{\temp}}{\widehat{\temp} + \widetilde{\temp}}
     \nabla \widehat{\temp}
  \right\}
  \,
  \cvolumee
  \\
  =
  \int_{\Omega}
  \frac{\kappa}{\cheatvol}
  \vectordot{\nabla \widehat{\temp}}{\nabla \widetilde{\entropy}}
  \,
  \cvolumee
  +
  \int_{\partial \Omega}
  \vectordot{
    \left\{
      \kappa 
      \left(
        1
        -
        \frac{\widehat{\temp}}{\widehat{\temp} + \widetilde{\temp}}
      \right)
      \nabla \widetilde{\temp}
      +
      \kappa 
      \frac{\widetilde{\temp}}{\widehat{\temp} + \widetilde{\temp}}
      \nabla \widehat{\temp}
    \right\}
  }
  {
    \vec{n}
  }
  \,
  \csurfacee
  =
  \int_{\Omega}
  \frac{\kappa}{\cheatvol}
  \vectordot{\nabla \widehat{\temp}}{\nabla \widetilde{\entropy}}
  \,
  \cvolumee
  ,
\end{multline}
where we have used the Stokes theorem and the fact that the perturbation $\widetilde{\temp}$ vanishes on the boundary. 

Further, we see that
\begin{equation}
  \label{eq:113}
  \int_{\Omega}
  \frac{\kappa}{\cheatvol}
  \vectordot{\nabla \widehat{\temp}}{\nabla \widetilde{\entropy}}
  \,
  \cvolumee
  =
  \int_{\Omega}
  \divergence
  \left\{
    \frac{\kappa}{\cheatvol}
    \widetilde{\entropy}
    \nabla \widehat{\temp}
  \right\}
  \,
  \cvolumee
  -
  \int_{\Omega}
  \divergence
  \left\{
    \frac{\kappa}{\cheatvol}
    \nabla \widehat{\temp} 
  \right\}
  \widetilde{\entropy}
  \,
  \cvolumee
  =
  0
  .
\end{equation}
Indeed, the first term vanishes in virtue of the Stokes theorem and the fact that $\widetilde{\temp}$ and consequently also $\widetilde{\entropy}$ vanish on the boundary. The second term vanishes in virtue of~\eqref{eq:non-equilibrium-rest-state} that holds for the non-equilibrium steady state~$\widehat{\temp}$. Consequently, we see that
\begin{equation}
  \label{eq:114}
  -
  \int_{\Omega}
  \widehat{\temp}
  \divergence
  \left(
    \frac{\kappa}{\cheatvol}
    \nabla \widetilde{\entropy}
  \right)
  \,
  \cvolumee
  +
  \int_{\Omega}
  \divergence \left( \kappa \nabla \widetilde{\temp} \right)
  \,
  \cvolumee
  =
  0,
\end{equation}
which means that~\eqref{eq:109} simplifies to
\begin{equation}
  \label{eq:115}
  \dd{}{t}
  \mathcal{V}_{\mathrm{neq}}
  \left(
    \left.
      \widetilde{\vec{W}}
    \right\|
    \widehat{\vec{W}}
  \right)
  =
  -
  \int_{\Omega}
  \frac{\kappa}{\cheatvol^2}
  \widehat{\temp} \vectordot{\nabla \widetilde{\entropy}}{\nabla \widetilde{\entropy}}
  \,
  \cvolumee
  -
  \int_{\Omega}
  \frac{2 \kappa}{\cheatvol^2}
  \widehat{\temp} \vectordot{\nabla \widetilde{\entropy}}{\nabla \widehat{\entropy}}
  \,
  \cvolumee
  .
\end{equation}

Let us now focus on the last term in~\eqref{eq:115}. We know that
\begin{equation}
  \label{eq:116}
  \nabla \widehat{\entropy}
  =
  \frac{\cheatvol}{\widehat{\temp}}
  \nabla \widehat{\temp},
\end{equation}
hence the term can be rewritten as
\begin{equation}
  \label{eq:117}
  \int_{\Omega}
  \frac{2 \kappa}{\cheatvol^2}
  \widehat{\temp} \vectordot{\nabla \widetilde{\entropy}}{\nabla \widehat{\entropy}}
  \,
  \cvolumee
  =
  \int_{\Omega}
  \frac{2 \kappa}{\cheatvol}
  \vectordot{\nabla \widetilde{\entropy}}{\nabla \widehat{\temp}}
  \,
  \cvolumee
  ,
\end{equation}
and using the same manipulation as in~\eqref{eq:113} we conclude that the term vanishes.

Finally, we see that
\begin{equation}
  \label{eq:118}
    \dd{}{t}
  \mathcal{V}_{\mathrm{neq}}
  \left(
    \left.
      \widetilde{\vec{W}}
    \right\|
    \widehat{\vec{W}}
  \right)
  =
  -
  \int_{\Omega}
  \frac{\kappa}{\cheatvol^2}
  \widehat{\temp} \vectordot{\nabla \widetilde{\entropy}}{\nabla \widetilde{\entropy}}
  \,
  \cvolumee
  ,
\end{equation}
hence the time derivative of
$
\mathcal{V}_{\mathrm{neq}}
  \left(
    \left.
      \widetilde{\vec{W}}
    \right\|
    \widehat{\vec{W}}
  \right)
$
is negative unless $\widetilde{\entropy}$ is equal to zero everywhere in $\Omega$. This means that~%
$
\mathcal{V}_{\mathrm{neq}}
  \left(
    \left.
      \widetilde{\vec{W}}
    \right\|
    \widehat{\vec{W}}
  \right)
$
is indeed a Lyapunov functional suitable for the nonlinear stability analysis of the steady non-equilibrium temperature field~$\widehat{\temp}$. Consequently, we see that the steady non-equilibrium state $\widehat{\temp}$ is unconditionally asymptotically stable.

\subsubsection{Relation to the standard energy method}
\label{sec:relat-class-energy-1}
We can again note that if the temperature perturbation $\widetilde{\temp}$ is small in the sense that $\frac{\widetilde{\temp}}{\widehat{\temp}} << 1$, then
\begin{equation}
  \label{eq:119}
  \mathcal{V}_{\mathrm{neq}}
  \left(
    \left.
      \widetilde{\vec{W}}
    \right\|
    \widehat{\vec{W}}
  \right)
  =
  \int_{\Omega}
  \rho
  \cheatvol
  \widehat{\temp}
  \left[
    \frac{\widetilde{\temp}}{\widehat{\temp}}
    -
    \ln \left( 1 + \frac{\widetilde{\temp}}{\widehat{\temp}} \right) 
  \right]
  \,
  \cvolumee
  \approx
  \int_{\Omega}
  \rho
  \cheatvol
  \frac{{\widetilde{\temp}}^2}{\widehat{\temp}}
  \,
  \cvolumee
  ,
\end{equation}
hence $\mathcal{V}_{\mathrm{neq}}$ is almost equal to the (square of) the weighted $\sleb{2}{\Omega}$ norm of the perturbation temperature field $\widetilde{\temp}$. Moreover, if~$\widehat{\temp}$ is position independent, that is if we analyse the \emph{equilibrium} rest state, we recover, up to a constant, the standard \emph{energy method} functional~$\mathcal{V}_{\mathrm{std}}$, see~\eqref{eq:21}.

Further, we see that the integrand in the standard Lyapunov functional $\mathcal{V}_{\mathrm{std}}$ is insensitive to the direction of the deviation from the non-equilibrium rest state. The integrand takes the same value both for $\widetilde{\temp}$ and $-\widetilde{\temp}$. On the other hand, the integrand in the Lyapunov functional~\eqref{eq:96} does not have this property. Its value is different for $-\widetilde{\temp}$ and $\widetilde{\temp}$, and, moreover, its value approaches infinity as $\widetilde{\temp} \to - \widehat{\temp}$.

One can also note that the \emph{relative entropy} functional, that is the functional ${\mathcal S}_{\mathrm{rel}} =_{\bydefinition} \int_{\Omega} \rho \widetilde{\entropy}\, \cvolumee$, see~\eqref{eq:98} and~\eqref{eq:99}, can not be used as a Lyapunov functional. It does not provide a control on the spatial distribution of the perturbations. In particular, ${\mathcal S}_{\mathrm{rel}} = 0$ does not imply that $\widetilde{\temp} = 0$ in the whole domain $\Omega$.

\subsubsection{Weak--strong uniqueness property}
\label{sec:stability-}
The notion of stability can be also understood as ``continuous dependence of thermodynamical processes upon initial state and supply terms'', see~\cite{dafermos.cm:second}, which is a different concept than that we have discussed above. In particular, the aim of the stability analysis understood in this sense is to show that if two solutions to a given initial-boundary value problem \emph{share the same initial condition and boundary condition}, then they coincide also for later times. This is a nontrivial question when the two solutions sharing the same initial conditions are for example the strong and the weak solution. Interestingly, a similar construction of a ``distance'' functional as outlined above have been used on \emph{ad hoc} grounds by~\cite{feireisl.e.jin.b.j.ea:relative,feireisl.e.novotny.a:weak-strong} in their seminal analysis of weak--strong uniqueness property for the Navier--Stokes--Fourier system. (See also~\cite{feireisl.e.novotny.a:singular}.) Note however that the weak--strong uniqueness analysis by~\cite{feireisl.e.jin.b.j.ea:relative,feireisl.e.novotny.a:weak-strong} is again restricted to a \emph{thermodynamically isolated} system, and it has no implications for the nonlinear stability analysis in the sense we are using in the present contribution.

\section{Conclusion}
\label{sec:conclusion}
We have shown that a Lyapunov functional suitable for the nonlinear stability analysis of steady solutions to the heat conduction problem in a rigid body can be constructed with the use of thermodynamical concepts. In particular, the thermodynamical concepts have been shown to be useful even in the case of a nonlinear stability analysis of a \emph{thermodynamically open} system. 

The outlined construction of the physically motivated Lyapunov functional is rather superfluous given \emph{the simple setting we have been studying}. In the present case, the standard \emph{energy method} definitely provides a formally much simpler approach to the nonlinear stability analysis. The construction however shows that nonlinear stability analysis can be indeed based on an insight into the physics behind the given system of governing equations. In particular, it indicates that using the square of the Lebesgue space norm of the temperature field as a Lyapunov functional is just a matter of mathematical convenience. The physically motivated Lyapunov functional is different from the mathematically convenient one.

More importantly, the outlined construction of a physically well-motivated Lyapunov functional seems to be \emph{general enough to be applied even in a more complex thermomechanical settings}. (Here, however, one can not in general expect unconditional stability, the non-equilibrium steady state must be expected to be stable only for some parameter values/size of the initial perturbation and so forth.) Indeed, the construction of the Lyapunov functional is in fact based only on the knowledge of the specific Helmholtz free energy $\fenergy$, which is known for many complex materials such as polymeric fluids, see for example \cite{dressler.m.edwards.bj.ea:macroscopic}, \cite{rajagopal.kr.srinivasa.ar:thermodynamic}, \cite{malek.j.rajagopal.kr.ea:on} and \cite{hron.j.milos.v.ea:on}. In such complicated settings the apparent complexity of the outlined construction could turn into an advantage, since the Lebesgue norm $\norm[\sleb{2}{\Omega}]{\cdot}$ used in the standard \emph{energy method} does not respect the natural physical background of the corresponding governing equations. Consequently, the advocated approach could provide a tool for nonlinear stability analysis in complex thermomechanical systems for which the standard \emph{energy method} has been so far unsuccessful.





\bibliographystyle{chicago}
\bibliography{vit-prusa}

\newpage
\appendix

\section{Example of stability analysis of a steady non-equilibrium state in a thermodynamically open system governed by a nonlinear equation}
\label{sec:stab-analys-steady}

Let us now document the use of the proposed method in a nonlinear setting. (Meaning that we are interested in the stability of a steady non-equilibrium state in a system described by a \emph{nonlinear governing equation}.) We again investigate heat conduction in a rigid body, but now the thermal conductivity is assumed to be a function of temperature.

We first reiterate on some concepts used in Section~\ref{sec:uncond-asympt-stab}, and we reformulate them in a form convenient for stability analysis of nonlinear heat conduction in a rigid body, see~Section~\ref{sec:reth-form-lyap}. The stability of a steady non-equilibrium solution to the nonlinear heat conduction problem is then analysed in Section~\ref{sec:heat-conduction-body}.

\subsection{Rethinking the formula for the Lyapunov functional and the time derivative of the Lyapunov functional}
\label{sec:reth-form-lyap}

\subsubsection{Candidate for Lyapunov functional in terms of specific Helmholtz free energy and its derivatives}
\label{sec:form-lyap-funct}

The candidate for Lyapunov functional is given by the formula~\eqref{eq:81}, that is
\begin{equation}
  \label{eq:120}
  \mathcal{V}_{\mathrm{neq}}
  \left(
    \left.
      \widetilde{\vec{W}}
    \right\|
    \widehat{\vec{W}}
  \right)
  =
  -
  \left\{
    \entropyrellp_{\widehat{\temp}}(\left. \widetilde{\vec{W}} \right\| \widehat{\vec{W}})
    -
    \energyrellp  (\left. \widetilde{\vec{W}} \right\| \widehat{\vec{W}})
  \right\}
  .
\end{equation}
If one needs to find an explicit formula for the Lyapunov functional, then one first needs to find a formula for the specific entropy $\entropy$ and the specific internal energy $\ienergy$. Since both these quantities can be expressed in terms of the specific Helmholtz free energy~$\fenergy$, it would be convenient to \emph{express the formula for the Lyapunov functional exclusively in terms of the specific Helmholtz free energy $\fenergy$ and its derivatives}. If we restrict ourselves to the setting where the specific Helmholtz free energy is a function of temperature only, and where the velocity field $\vec{v}$ vanishes\footnote{Again, all the computations are easy to generalise to a more complex setting.}, then the only state variable is the temperature, $\vec{W} = \temp$, and we can proceed as follows.

Functionals $\entropyrellp_{\widehat{\temp}}(\left. \widetilde{\vec{W}} \right\| \widehat{\vec{W}})$ and $\energyrellp(\left. \widetilde{\vec{W}} \right\| \widehat{\vec{W}})$, see~\eqref{eq:82}, reduce to
\begin{subequations}
  \label{eq:121}
  \begin{align}
    \label{eq:122}
    \entropyrellp_{\widehat{\temp}}(\left. \widetilde{\vec{W}} \right\| \widehat{\vec{W}})
    &=
    \int_{\Omega} 
    \left(
      \rho \widehat{\temp} \entropy(\widehat{\temp} + \widetilde{\temp})
      -
      \rho \widehat{\temp} \entropy(\widehat{\temp})
      -
      \rho \widehat{\temp} \left. \pd{\entropy}{\temp} \right|_{\temp = \widehat{\temp}} \widetilde{\temp}
    \right)
    \,
    \cvolumee
    ,
    \\
    \label{eq:123}
    \energyrellp(\left. \widetilde{\vec{W}} \right\| \widehat{\vec{W}})
    &=
    \int_{\Omega} 
    \left(
      \rho \ienergy(\widehat{\temp} + \widetilde{\temp})
      -
      \rho \ienergy(\widehat{\temp})
      -
      \rho \left. \pd{\ienergy}{\temp} \right|_{\temp = \widehat{\temp}} \widetilde{\temp}
    \right)
    \,
    \cvolumee
    ,
  \end{align}
\end{subequations}
which yields
\begin{equation}
  \label{eq:150}
  \mathcal{V}_{\mathrm{neq}}
  \left(
    \left.
      \widetilde{\vec{W}}
    \right\|
    \widehat{\vec{W}}
  \right)
  =
  -
  \left\{
    \entropyrellp_{\widehat{\temp}}(\left. \widetilde{\vec{W}} \right\| \widehat{\vec{W}})
    -
    \energyrellp  (\left. \widetilde{\vec{W}} \right\| \widehat{\vec{W}})
  \right\}
  =
  -
  \int_{\Omega} 
  \rho
  \left[
    \widehat{\temp} \entropy(\widehat{\temp} + \widetilde{\temp})
    -
    \widehat{\temp} \entropy(\widehat{\temp})
    -
    \widehat{\temp} \left. \pd{\entropy}{\temp} \right|_{\temp = \widehat{\temp}} \widetilde{\temp}
    -
    \ienergy(\widehat{\temp} + \widetilde{\temp})
    +
    \ienergy(\widehat{\temp})
    +
    \left. \pd{\ienergy}{\temp} \right|_{\temp = \widehat{\temp}} \widetilde{\temp}
  \right]
  \,
  \cvolumee
  .
\end{equation}
The specific Helmholtz free energy $\fenergy$ is given as the Legendre transform of the internal energy 
\begin{equation}
  \label{eq:151}
  \fenergy = \ienergy - \temp \entropy,
\end{equation}
the entropy $\entropy$ is obtained from the Helmholtz free energy $\fenergy$ via~\eqref{eq:23}, that is
\begin{equation}
  \label{eq:176}
  \entropy = - \pd{\fenergy}{\temp}.
\end{equation}
and the derivative of the internal energy~$\ienergy$ with respect to the temperature~$\temp$ can be equivalently expressed as
\begin{equation}
  \label{eq:149}
  -\temp \ppd{\fenergy}{\temp} = \pd{\ienergy}{\temp}.
\end{equation}
Using~\eqref{eq:151}, \eqref{eq:176} and~\eqref{eq:149} in \eqref{eq:150} gives us
\begin{equation}
  \label{eq:152}
  \mathcal{V}_{\mathrm{neq}}
  \left(
    \left.
      \widetilde{\vec{W}}
    \right\|
    \widehat{\vec{W}}
  \right)
  =
  -
  \int_{\Omega} 
  \rho
  \left[
    -
    \left\{
      \ienergy(\widehat{\temp} + \widetilde{\temp})
      -
      \widehat{\temp} \entropy(\widehat{\temp} + \widetilde{\temp})
      \vphantom{\frac{1}{2}}
    \right\}
    +
    \left\{
      \ienergy(\widehat{\temp})
      -
      \widehat{\temp} \entropy(\widehat{\temp})
      \vphantom{\frac{1}{2}}
    \right\}
  \right]
  \,
  \cvolumee
  =
  -
  \int_{\Omega} 
  \rho
  \left[
    -
    \fenergy(\widehat{\temp} + \widetilde{\temp})
    +
    \fenergy(\widehat{\temp})
    -
    \widetilde{\temp}
    \entropy(\widehat{\temp} + \widetilde{\temp})
  \right]
  \,
  \cvolumee
  ,
\end{equation}
which can be rewritten as
\begin{subequations}
  \label{eq:153}
  \begin{equation}
    \mathcal{V}_{\mathrm{neq}}
    \left(
      \left.
        \widetilde{\vec{W}}
      \right\|
      \widehat{\vec{W}}
    \right)
    =
    \int_{\Omega} 
    \rho
    \Psi(\widehat{\temp}, \widetilde{\temp})
    \,
    \cvolumee
    ,
\end{equation}
where
\begin{equation}
  \label{eq:178}
  \Psi(\widehat{\temp}, \widetilde{\temp})
  =_{\bydefinition}
  \fenergy(\widehat{\temp} + \widetilde{\temp})
  -
  \fenergy(\widehat{\temp})
  -
  \widetilde{\temp}
  \left. \pd{\fenergy}{\temp} \right|_{\temp = \widehat{\temp} + \widetilde{\temp}}
  .
\end{equation}
\end{subequations}
This is the sought formula for the candidate for Lyapunov functional in terms of the specific Helmholtz free energy $\fenergy$.  Note that the last term in~\eqref{eq:178} contains the derivative of $\fenergy$ evaluated at $\temp = \widehat{\temp} + \widetilde{\temp}$. This means that the terms
$
-
\fenergy(\widehat{\temp})
-
\left. \pd{\fenergy}{\temp} \right|_{\temp = \widehat{\temp} + \widetilde{\temp}} \widetilde{\temp}
$
\emph{are not equal} to the first two terms of Taylor expansion of 
$
-\fenergy(\widehat{\temp} + \widetilde{\temp})
$
which read
$
-
\fenergy(\widehat{\temp})
-
\left. \pd{\fenergy}{\temp} \right|_{\temp = \widehat{\temp}}
\widetilde{\temp}
$%
.

\subsubsection{Time derivative of the Lyapunov functional}
\label{sec:time-deriv-lyap-1}
If we assume that the energy flux $\efluxc$ is given by a linear constitutive relation
\begin{equation}
  \label{eq:155}
  \efluxc = - \kapparef \nabla \temp,
\end{equation}
where $\kapparef$ is a constant, then a close inspection of the calculations done in Section~\ref{sec:time-deriv-prop} reveals that the time derivative of the candidate for Lyapunov functional~\eqref{eq:81}, and hence of the functional~\eqref{eq:153}, is nonpositive \emph{irrespective of the particular choice of the specific Helmholtz free energy}.

Indeed, the candidate for Lyapunov functional is given by the formula
\begin{equation}
  \label{eq:156}
  \mathcal{V}_{\mathrm{neq}}
  \left(
    \left.
      \widetilde{\vec{W}}
    \right\|
    \widehat{\vec{W}}
  \right)
  =
  -
  \int_{\Omega} 
  \rho
  \widehat{\temp} 
  \left[
    \entropy(\widehat{\temp} + \widetilde{\temp})
    -
    \entropy(\widehat{\temp})
  \right]
  \,
  \cvolumee
  +
  \int_{\Omega} 
  \rho
  \ienergy(\widehat{\temp} + \widetilde{\temp})
  \,
  \cvolumee
  -
  \int_{\Omega} 
  \rho
  \ienergy(\widehat{\temp})
  \,
  \cvolumee
  ,
\end{equation}
see~\eqref{eq:152}. If the energy flux $\efluxc$ is given by the formula~\eqref{eq:155}, and if we deal with heat conduction in a rigid body, then the evolution equations for the specific entropy $\entropy$ and the internal energy $\ienergy$ read 
\begin{subequations}
  \label{eq:154}
  \begin{align}
    \label{eq:157}
    \rho \pd{\ienergy}{t} &= \divergence \left( \kapparef \nabla \temp \right), \\
    \label{eq:158}
    \rho
    \pd{\entropy}{t}
    &=
    \kapparef\frac{\vectordot{\nabla \temp}{\nabla \temp}}{\temp^2}
    +
    \divergence
    \left(
      \kapparef \frac{\nabla \temp}{\temp}
    \right)
    ,
  \end{align}
\end{subequations}
see \eqref{eq:30} and~\eqref{eq:31}. Using~\eqref{eq:154} in taking the time derivative of~\eqref{eq:156} yields
\begin{multline}
  \label{eq:159}
  \dd{}{t}
  \mathcal{V}_{\mathrm{neq}}
  \left(
    \left.
      \widetilde{\vec{W}}
    \right\|
    \widehat{\vec{W}}
  \right)
  =
  -
  \int_{\Omega} 
  \rho
  \widehat{\temp} 
  \pd{}{t}
  \left[
    \entropy(\widehat{\temp} + \widetilde{\temp})
    -
    \entropy(\widehat{\temp})
  \right]
  \,
  \cvolumee
  +
  \int_{\Omega} 
  \rho
  \pd{\ienergy(\widehat{\temp} + \widetilde{\temp})}{t}
  \,
  \cvolumee
  -
  \int_{\Omega} 
  \rho
  \pd{\ienergy(\widehat{\temp})}{t}
  \,
  \cvolumee
  \\
  =
  -
  \int_{\Omega} 
  \widehat{\temp} 
  \left[
    \kapparef
    \vectordot{\nabla \Theta (\widehat{\temp} + \widetilde{\temp})}{\nabla \Theta (\widehat{\temp} + \widetilde{\temp})}
    -
    \kapparef
    \vectordot{\nabla \Theta (\widehat{\temp})}{\nabla \Theta (\widehat{\temp})}
    +
    \divergence
    \left(
      \kapparef
      \nabla \Theta(\widehat{\temp} + \widetilde{\temp})
    \right)
    -
    \divergence
    \left(
      \kapparef
      \nabla \Theta(\widehat{\temp})
    \right)
    \vphantom{\frac{1}{2}}
  \right]
  \,
  \cvolumee
  \\
  +
  \int_{\Omega} 
  \divergence \left(\kapparef \nabla \widetilde{\temp} \right)
  \,
  \cvolumee
  ,
\end{multline}
where we have denoted
\begin{equation}
  \label{eq:160}
  \Theta (\temp) =_{\bydefinition} \ln \left( \frac{\temp}{\tempref} \right).
\end{equation}
If we further introduce the notation
\begin{subequations}
  \label{eq:162}
  \begin{align}
    \label{eq:163}
    \widehat{\Theta} &= _{\bydefinition} \ln \left( \frac{\widehat{\temp}}{\tempref} \right), \\
    \label{eq:164}
    \widetilde{\Theta} &=_{\bydefinition} \ln \left( 1 + \frac{\widetilde{\temp}}{\widehat{\temp}} \right)
    ,
  \end{align}
\end{subequations}
we see that~\eqref{eq:159} can be rewritten as
\begin{equation}
  \label{eq:161}
  \dd{}{t}
  \mathcal{V}_{\mathrm{neq}}
  \left(
    \left.
      \widetilde{\vec{W}}
    \right\|
    \widehat{\vec{W}}
  \right)
  =
  -
  \int_{\Omega} 
  \widehat{\temp} 
  \left[
    \kapparef
    \vectordot{\nabla \widetilde{\Theta}}{\nabla \widetilde{\Theta}}
    +
    2
    \kapparef
    \vectordot{\nabla \widehat{\Theta}}{\nabla \widetilde{\Theta}}
    +
    \divergence
    \left(
      \kapparef
      \nabla \widetilde{\Theta}
    \right)
    \vphantom{\frac{1}{2}}
  \right]
  \,
  \cvolumee
  +
  \int_{\Omega} 
  \divergence \left(\kapparef \nabla \widetilde{\temp} \right)
  \,
  \cvolumee
  .
\end{equation}

The only difference between~\eqref{eq:161} and \eqref{eq:109} is that the function $\Theta(\temp)$ is not necessarily directly related to the entropy as in the case studied in Section~\ref{sec:time-deriv-prop}. (Recall that in the latter case the entropy is given by the formula~\eqref{eq:24}, hence~$\widetilde{\entropy} = \cheatvol \widetilde{\Theta}$.) Function $\Theta(\temp)$ is just an auxiliary function, that allows us to manipulate the right-hand side of~\eqref{eq:161} exactly in the same manner as in Section~\ref{sec:time-deriv-prop}. Repeating all the calculations discussed in Section~\ref{sec:time-deriv-prop}, we arrive to the conclusion that the time derivative of the candidate for Lyapunov functional is given by the formula
\begin{equation}
  \label{eq:165}
  \dd{}{t}
  \mathcal{V}_{\mathrm{neq}}
  \left(
    \left.
      \widetilde{\vec{W}}
    \right\|
    \widehat{\vec{W}}
  \right)
  =
  -
  \int_{\Omega} 
  \kapparef
  \widehat{\temp} 
  \vectordot{\nabla \widetilde{\Theta}}{\nabla \widetilde{\Theta}}
  \,
  \cvolumee
  .
\end{equation}
The derivative is nonpositive, and it vanishes if and only if $\widetilde{\Theta}$ vanishes in $\Omega$. 

This observation shows that the time derivative of the Lyapunov functional does not depend on a particular formula for the \emph{entropy}. (And since the formula for the entropy $\entropy$ is a consequence of the specification of the Helmholtz free energy $\fenergy$, the time derivative of the functional does not depend on the specific choice of the Helmholtz free energy.) The only thing that in the present case matters regarding the time derivative of 
$
\mathcal{V}_{\mathrm{neq}}
  \left(
    \left.
      \widetilde{\vec{W}}
    \right\|
    \widehat{\vec{W}}
  \right)
$ 
is the formula for the energy flux $\efluxc$ that is closely related to the entropy production. Such an observation is perfectly reasonable. \emph{The construction of Lyapunov functional in principle reflects \emph{energy storage mechanisms} in the body of interest, while its time derivative is in principle determined by the \emph{entropy production mechanisms}, and these two mechanisms are considered to be independent.}

\subsection{Stability analysis of heat conduction in a rigid body with a temperature dependent thermal conductivity}
\label{sec:heat-conduction-body}

\subsubsection{Nonlinear heat conduction equation}
\label{sec:nonl-heat-cond}
We consider heat conduction in a rigid body, but unlike in Section~\ref{sec:general-steady-state} the thermal conductivity $\kappa$ is now assumed to be a function of temperature. In particular, we assume that
\begin{equation}
  \label{eq:124}
    \kappa(\temp) =_{\bydefinition} \kappa_{\reference} f(\temp), 
    \qquad
    f(\temp) =_{\bydefinition} \exponential{\alpha \frac{\temp - \temp_{\reference}}{\temp_{\reference}}}
    ,
\end{equation}
where $\kappa_{\reference}$ is a positive constant,  $\alpha$ is a nonegative constant a $\temp_{\reference}$ is a reference temperature. This choice is convenient since it leads to explicit formulae for all involved quantities, especially for the Lyapunov functional constructed by the method outlined above. Note, however, that the particular formula for $\kappa(\temp)$ is not too much important, the important fact is the monotonicity of~$\kappa(\temp)$.

The evolution of the temperature field $\temp$ in the domain $\Omega$ is described by the equation
\begin{subequations}
  \label{eq:127}
  \begin{align}
    \label{eq:125}
    \rho {\cheatvolref} \pd{\temp}{t} &= \divergence \left( \kappa(\temp) \nabla \temp \right), \\
    \label{eq:126}
    \left. \temp \right|_{\partial \Omega} &= \tempbdr, \\
    \label{eq:128}
    \left. \temp \right|_{t=0} &= \tempinit,
  \end{align}
\end{subequations}
where $\tempbdr$ denotes the temperature value on the boundary, and $\tempinit$ denotes the initial temperature distribution. Symbol~$\cheatvolref$ denotes the specific heat capacity at constant volume, which is assumed to be a positive constant. We assume that $\widehat{\temp}$ is a steady non-equilibrium solution to~\eqref{eq:127}, that is $\widehat{\temp}$ solves the boundary value problem
\begin{subequations}
  \label{eq:131}
  \begin{align}
    \label{eq:129}
    0 &= \divergence \left( \kappa(\widehat{\temp}) \nabla \widehat{\temp} \right), \\
    \label{eq:130}
    \left. \widehat{\temp} \right|_{\partial \Omega} &= \tempbdr.
  \end{align}
\end{subequations}
As before, we are interested in the stability of this steady non-equilibrium state, and we want to characterise its stability using a Lyapunov functional.

\subsubsection{Formulation of an auxiliary problem -- temperature dependent thermal conductivity versus temperature dependent specific heat capacity}
\label{sec:temp-depend-therm}

We note that \eqref{eq:131} can be rewritten as
\begin{equation}
  \label{eq:134}
   0 = \divergence \left( \kappa_{\reference} \nabla F(\temp) \right),
\end{equation}
where
\begin{equation}
  \label{eq:135}
  F(\temp) =_{\bydefinition} \int_{s = 0}^{\temp} f(s) \, \diff s = \frac{\tempref}{\alpha} \left( \exponential{\alpha \frac{\temp - \temp_{\reference}}{\temp_{\reference}}} - \exponential{-\alpha}\right)
\end{equation}
denotes a primitive function to the function $f$ introduced in~\eqref{eq:124}. Since $\kappa (\temp)$ is a strictly increasing function, we can introduce a new variable $\vartemp$, \emph{or a new temperature scale}, via 
\begin{equation}
  \label{eq:167}
  \vartemp =_{\bydefinition}  F(\temp).
\end{equation}
(Note that both temperature scales coincide as $\alpha \to 0+$, that is $\vartemp \to \temp$ as $\alpha \to 0+$.) Using the newly introduced rescaled temperature $\vartemp$, we can rewrite~\eqref{eq:134} as
\begin{equation}
  \label{eq:136}
   0 = \divergence \left( \kappa_{\reference} \nabla \vartemp \right).
\end{equation}
Further, rewriting the evolution equation~\eqref{eq:127} in terms of $\vartemp$ yields
\begin{equation}
  \label{eq:137}
  \rho {\cheatvolref} \dd{\inverse{F}}{\vartemp} \pd{\vartemp}{t} = \divergence \left( \kappa_{\reference} \nabla \vartemp \right),
\end{equation}
where $\inverse{F}$ denotes the inverse to the function $F$. This equation can be interpreted as a heat conduction equation for a material with a \emph{temperature dependent specific heat capacity at constant volume} $\cheatvol$ and a \emph{linear} constitutive relation for the energy flux $\efluxc$. 

Indeed, if the energy flux $\efluxc$ is a linear function of the gradient of temperature, $\efluxc = - \kappa_{\reference} \nabla \vartemp$, then the general evolution equation for the temperature reads
\begin{equation}
  \label{eq:138}
  \rho \cheatvol(\vartemp) \pd{\vartemp}{t} =  \divergence \left( \kappa_{\reference} \nabla \vartemp \right),
\end{equation}
where the specific heat capacity at constant volume $\cheatvol$ is given by the formula 
\begin{equation}
  \label{eq:139}
  \cheatvol(\vartemp) =_{\bydefinition} - \vartemp \ppd{\fenergy}{\vartemp}
\end{equation}
and $\fenergy$ denotes the specific Helmholtz free energy. Consequently, \eqref{eq:138} is tantamount to~\eqref{eq:137} provided that we define the specific Helmholtz free energy appropriately. 

Problem~\eqref{eq:127} with temperature dependent thermal conductivity $\kappa$ and constant specific heat capacity $\cheatvolref$ is therefore formally equivalent to the problem~\eqref{eq:138} for a material with temperature dependent specific heat capacity 
\begin{equation}
  \label{eq:140}
  \cheatvol(\vartemp) = {\cheatvolref} \dd{\inverse{F}}{\vartemp}
\end{equation}
and constant thermal conductivity $\kappa_{\reference}$. This reformulation of the original problem turns out to be more suitable for the ongoing stability analysis.

Concerning equation~\eqref{eq:138} with temperature dependent $\cheatvol$, we get very close to the setting discussed in Section~\ref{sec:constr-phys-motiv-1} and Section~\ref{sec:time-deriv-prop}. In particular, we know how to construct a physically motivated candidate for Lyapunov functional, see Section~\ref{sec:constr-lyap-funct} and Appendix~\ref{sec:reth-form-lyap}. All we need to do is to identify the formula for the Helmholtz free energy $\fenergy$ from the equation
\begin{equation}
  \label{eq:141}
   - \vartemp \ppd{\fenergy}{\vartemp} = \cheatvolref \dd{\inverse{F}}{\vartemp},
\end{equation}
and substitute for $\fenergy$ in~\eqref{eq:153}. Further, the linearity of the energy flux $\efluxc = - \kappa_{\reference} \nabla \vartemp$ brings us to the setting discussed in~Section~\ref{sec:time-deriv-lyap-1}. Consequently, the time derivative of the candidate for Lyapunov functional is nonpositive, and vanishes if and only if the perturbation field $\widetilde{\vartemp}$ vanishes.  

It remains to verify that the candidate for the Lyapunov functional is nonnegative and that it vanishes if and only if the perturbation field $\widetilde{\vartemp}$ vanishes. This will complete the stability analysis for the auxiliary problem~\eqref{eq:138}, and consequently also for the original problem~\eqref{eq:127}. Let us now proceed with the outlined plan.

\subsubsection{Identification of specific Helmholtz free energy}
\label{sec:ident-spec-helmh}
Evaluating the derivative of $\inverse{F}$ yields
\begin{equation}
  \label{eq:142}
  \dd{\inverse{F}}{\vartemp}
  =
  \frac{1}{\left. \dd{F}{\temp} \right|_{\temp = \inverse{F}(\vartemp)}}
  =
  \frac{1}{\left. f(\temp) \right|_{\temp = \inverse{F}(\vartemp)}}
  =
  \frac{1}{\alpha \frac{\vartemp}{\tempref} + \exponential{-\alpha}},
\end{equation}
hence equation~\eqref{eq:141} for the specific Helmholtz free energy $\fenergy$ reads
\begin{equation}
  \label{eq:143}
  \ddd{\fenergy}{\vartemp} = - \frac{\cheatvolref}{\vartemp \left(\alpha \frac{\vartemp}{\tempref} + \exponential{-\alpha} \right)}.
\end{equation}
The integration leads to
\begin{equation}
  \label{eq:144}
  \dd{\fenergy}{\vartemp} 
  =
  -
  \cheatvolref
  \exponential{\alpha}
  \left[
    \ln
    \left(
      \frac{\vartemp}{\vartempref}
    \right)
    -
    \ln
    \left(
      \frac{\alpha \frac{\vartempref}{\tempref} \frac{\vartemp}{\vartempref} + \exponential{-\alpha}}{\alpha \frac{\vartempref}{\tempref} + \exponential{-\alpha}}
    \right)
  \right]
\end{equation}
where we have conveniently fixed the integration constant in such a way that the derivative $\dd{\fenergy}{\vartemp}$ vanishes at $\vartemp = \vartempref$. This means that we fix the zero entropy level at $\vartemp = \vartempref$. Further, we can chose $\vartempref$ to be equal to the rescaled temperature value $\vartemp$ that corresponds to $\tempref$, that is $\vartempref =_{\bydefinition} F(\tempref)$, which yields $\vartempref = \frac{\tempref}{\alpha} \left(1 - \exponential{-\alpha} \right)$. If we do so, then~\eqref{eq:144} reads\footnote{Note that when inspecting the limit $\alpha \to 0+$, we have to take into account that $\vartempref = \vartempref(\alpha)$ and so forth.}
\begin{equation}
  \label{eq:166}
  \dd{\fenergy}{\vartemp} 
  =
  -
  \cheatvolref
  \exponential{\alpha}
  \left[
    \ln
    \left(
      \frac{\vartemp}{\vartempref}
    \right)
    -
    \ln
    \left(
      \left(1 - \exponential{-\alpha} \right) \frac{\vartemp}{\vartempref} + \exponential{-\alpha}
    \right)
  \right]
  .
\end{equation}

Further integration yields
\begin{equation}
  \label{eq:145}
  \fenergy
  =
  -
  \cheatvolref
  \exponential{\alpha}
  \vartemp
  \left[
    \ln
    \left(
      \frac{\vartemp}{\vartempref}
    \right)
    -
    1
  \right]
  +
  \cheatvolref
  \exponential{\alpha}
  \vartemp
  \left[
    \ln
    \left(
          \left(1 - \exponential{-\alpha} \right) \frac{\vartemp}{\vartempref} + \exponential{-\alpha}
    \right)
    +
    \frac{
      \ln
      \left(
        \left(1 - \exponential{-\alpha} \right) \frac{\vartemp}{\vartempref} + \exponential{-\alpha}
      \right)
    }
    {
      \frac{1 - \exponential{-\alpha}}{\exponential{-\alpha}} \frac{\vartemp}{\vartempref}
    }
    -
    1
  \right]
  ,
\end{equation}
where we have set the integration constant equal to zero. If $\alpha$ tends to zero, $\alpha \to 0+$, we recover, up to an inconsequential additive constant, the standard formula~\eqref{eq:22} for the specific Helmholtz free energy in a material with a constant specific heat capacity at constant volume.

\subsubsection{Lyapunov functional for the auxiliary problem}
\label{sec:lyapunov-functional}

Having identified the specific Helmholtz free energy~$\fenergy$, we are ready to construct the candidate for Lyapunov functional. According to~\eqref{eq:153}, we need to evaluate the expression
\begin{equation}
  \label{eq:147}
  \Psi (\widehat{\vartemp}, \widetilde{\vartemp}) =_{\bydefinition} \fenergy(\widehat{\vartemp} + \widetilde{\vartemp}) - \fenergy(\widehat{\vartemp})  - \widetilde{\vartemp} \left. \pd{\fenergy}{\vartemp} \right|_{\vartemp = \widehat{\vartemp} + \widetilde{\vartemp}},
\end{equation}
which for $\fenergy$ given by~\eqref{eq:145} yields
\begin{equation}
  \label{eq:148}
  \Psi (\widehat{\vartemp}, \widetilde{\vartemp})
  =
  \cheatvolref
  \exponential{\alpha}
  \widehat{\vartemp}
  \left[
    \frac{\widetilde{\vartemp}}{\widehat{\vartemp}}
    -
    \ln \left( 1 + \frac{\widetilde{\vartemp}}{\widehat{\vartemp}} \right) 
  \right]
  +
  \cheatvolref
  \exponential{\alpha}
  \widehat{\vartemp}
  \left[
    \frac{\exponential{-\alpha} + \left( 1  - \exponential{-\alpha} \right) \frac{\widehat{\vartemp}}{\vartempref}}{\left( 1  - \exponential{-\alpha} \right) \frac{\widehat{\vartemp}}{\vartempref}}
    \ln
    \left(
      \frac{
        \exponential{-\alpha} + \left( 1  - \exponential{-\alpha} \right) \frac{\widehat{\vartemp}}{\vartempref} \left( 1 + \frac{\widetilde{\vartemp}}{\widehat{\vartemp}}\right)
      }
      {
        \exponential{-\alpha} + \left( 1  - \exponential{-\alpha} \right) \frac{\widehat{\vartemp}}{\vartempref}
      }
    \right)
    -
    \frac{\widetilde{\vartemp}}{\widehat{\vartemp}}
  \right]
  .
\end{equation}
Using this expression for $\Psi (\widehat{\vartemp}, \widetilde{\vartemp})$, it is straightforward to see that if $\alpha \to 0+$, then one recovers the expression used in the Lyapunov functional for heat conduction problem with constant specific heat capacity at constant volume, see~\eqref{eq:96}. 

The expression for $\Psi (\widehat{\vartemp}, \widetilde{\vartemp})$ can be further rewritten as
\begin{subequations}
  \label{eq:169}
  \begin{equation}
    \label{eq:168}
    \Psi (\widehat{\vartemp}, \widetilde{\vartemp})
    =
    \frac{
      \cheatvolref \widehat{\vartemp}
    }
    {
      b
    }
    \left[
      \frac{1}{a}
      \ln 
      \left(
        1 + a \frac{\widetilde{\vartemp}}{\widehat{\vartemp}}
      \right)
      -
      \ln \left(1 + \frac{\widetilde{\vartemp}}{\widehat{\vartemp}}  \right)
    \right]
    ,
  \end{equation}
  where
  \begin{equation}
    \label{eq:170}
    b =_{\bydefinition} \exponential{-\alpha}, \qquad
    a =_{\bydefinition} \frac{ 
      \left( 1 - \exponential{-\alpha} \right) \frac{\widehat{\vartemp}}{\vartempref} 
    }
    { 
      \exponential{-\alpha} 
      +
      \left( 1 - \exponential{-\alpha} \right) \frac{\widehat{\vartemp}}{\vartempref}
    }
    .
  \end{equation}
\end{subequations}

Let us now investigate the properties of function $\Psi (\widehat{\vartemp}, \widetilde{\vartemp})$. First, we see that the parameters $a$ and $b$ satisfy $b \leq 1$ and $0 \leq a < 1$, and that $b \to 1$ and $a \to 0$ as $\alpha \to 0+$. Second, we see that $\Psi (\widehat{\vartemp}, \widetilde{\vartemp})$ vanishes at the point $\widetilde{\vartemp} = 0$. 

\begin{figure}[h]
  \centering
  \subfloat[Large scale behaviour.]{\includegraphics[width=0.45\textwidth]{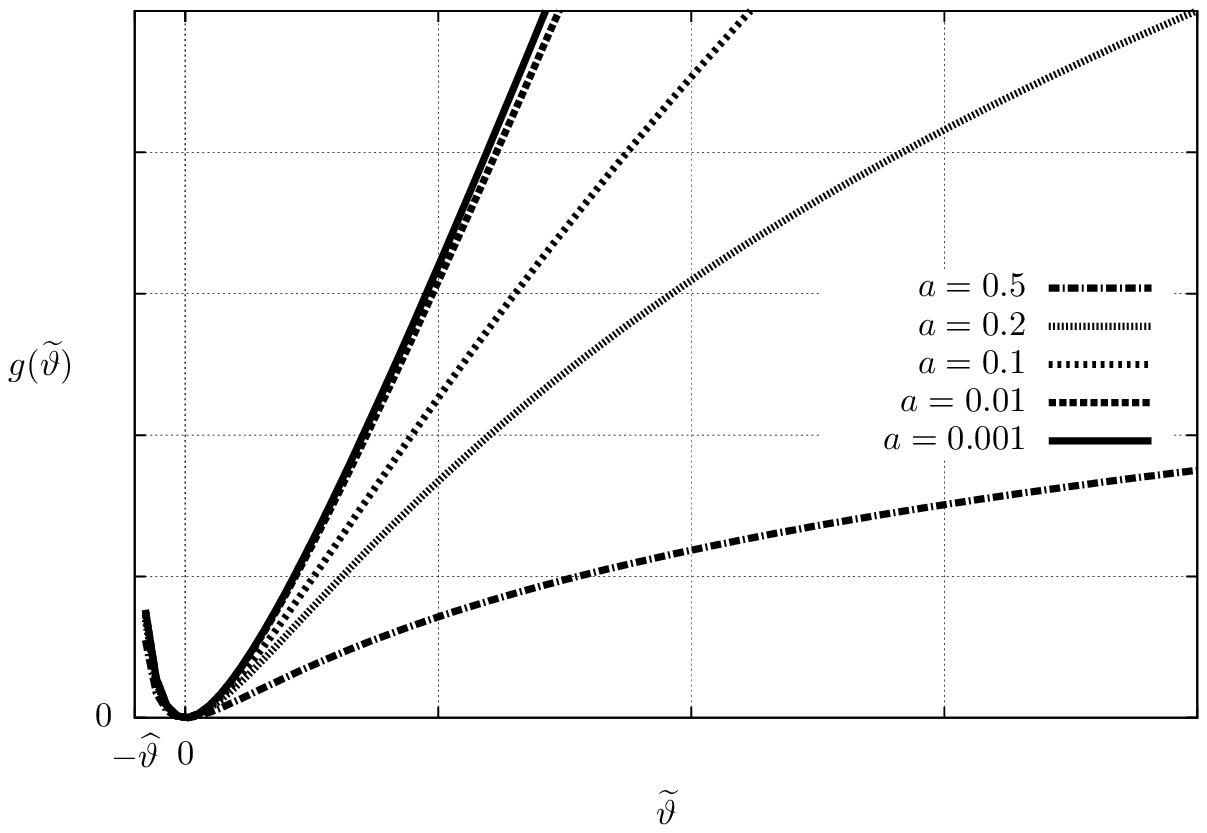}}
  \qquad
  \subfloat[Behaviour in the neighborhood of zero.]{\includegraphics[width=0.45\textwidth]{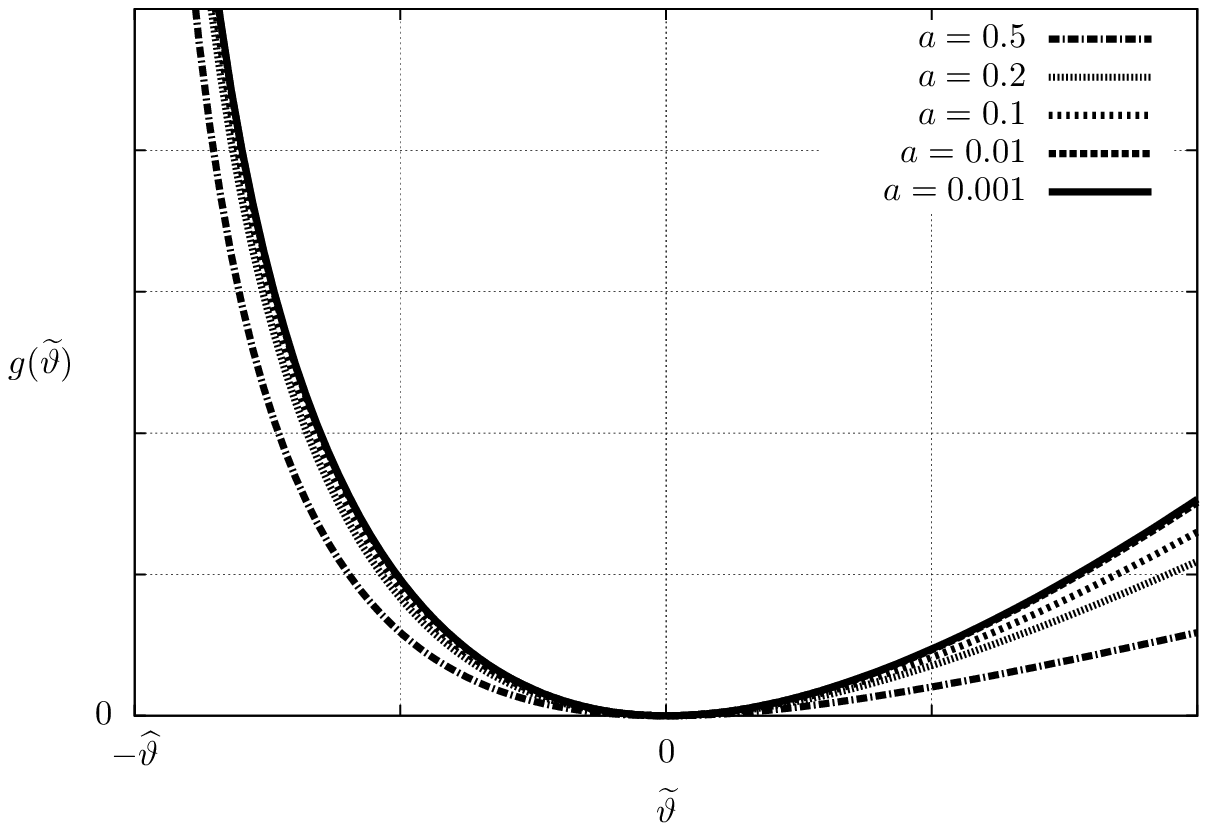}}
  \caption{Auxiliary function $g(\widetilde{\vartemp}) 
    =  
    \frac{1}{a}
    \ln 
    \left(
      1 + a \frac{\widetilde{\vartemp}}{\widehat{\vartemp}}
    \right)
    -
    \ln \left(1 + \frac{\widetilde{\vartemp}}{\widehat{\vartemp}}  \right)$ for various values of the parameter $a$.}
  \label{fig:function-g}
\end{figure}

Finally, a close inspection of function
\begin{equation}
  \label{eq:172}
  g(\widetilde{\vartemp}) 
  =_{\bydefinition}
  \frac{1}{a}
  \ln 
  \left(
    1 + a \frac{\widetilde{\vartemp}}{\widehat{\vartemp}}
  \right)
  -
  \ln \left(1 + \frac{\widetilde{\vartemp}}{\widehat{\vartemp}}  \right),
\end{equation}
that constitutes the key part of function $\Psi (\widehat{\vartemp}, \widetilde{\vartemp})$, reveals that this function is well defined provided that $\widetilde{\vartemp} > -\widehat{\vartemp}$. This means that $\Psi (\widehat{\vartemp}, \widetilde{\vartemp})$ is well defined whenever the complete temperature field $\vartemp = \widehat{\vartemp} + \widetilde{\vartemp}$ is positive, which is granted by the properties of the evolution equation for the temperature field. Moreover if $0 \leq a < 1$ and if $\widehat{\vartemp} > 0$, then the function $g(\widetilde{\temp})$ is a nonnegative function, and it vanishes if and only if $\widetilde{\temp} = 0$, see also Figure~\ref{fig:function-g}. (Note that function $g$ is not, for certain values of parameter $a$, a convex function. If $a=0$ we define $g(\widetilde{\vartemp})$ as the limit for $a \to 0+$.) This confirms that the functional
\begin{equation}
  \label{eq:146}
  \mathcal{V}_{\mathrm{neq}}
  \left(
    \left.
      \widetilde{\vec{W}}
    \right\|
    \widehat{\vec{W}}
  \right)
  =
  \int_{\Omega}
  \rho
  \Psi (\widehat{\vartemp}, \widetilde{\vartemp})
  \,
  \cvolumee
  =
  \int_{\Omega}
  \rho
  \frac{
    \cheatvolref \widehat{\vartemp}
  }
  {
    b
  }
  \left[
    \frac{1}{a}
    \ln 
    \left(
      1 + a \frac{\widetilde{\vartemp}}{\widehat{\vartemp}}
    \right)
    -
    \ln \left(1 + \frac{\widetilde{\vartemp}}{\widehat{\vartemp}}  \right)
  \right]
  \,
  \cvolumee
\end{equation}
is indeed a good candidate for Lyapunov functional for stability analysis of the auxiliary problem~\eqref{eq:138}. It remains to check whether the value of 
$\mathcal{V}_{\mathrm{neq}}
\left(
  \left.
    \widetilde{\vec{W}}
  \right\|
  \widehat{\vec{W}}
\right)$
decreases as $\vartemp = \widehat{\vartemp} + \widetilde{\vartemp}$ evolves in time according to~\eqref{eq:138}.

\subsubsection{Time derivative of Lyapunov functional for the auxiliary problem}
\label{sec:time-deriv-lyap}
The auxiliary problem~\eqref{eq:138} is a heat conduction problem in a rigid body with \emph{temperature dependent} specific heat capacity at constant volume. The energy flux is however given by a linear constitutive relation $\efluxc = - \kapparef \nabla \vartemp$, which is the setting we have discussed in Section~\ref{sec:time-deriv-lyap-1}. The formula for the time derivative therefore reads
\begin{subequations}
  \label{eq:177}
  \begin{equation}
  \label{eq:173}
  \dd{
  }
  {
    t
  }
  \mathcal{V}_{\mathrm{neq}}  
  \left(
    \left.
      \widetilde{\vec{W}}
    \right\|
    \widehat{\vec{W}}
  \right)
  =
  -
  \int_{\Omega} 
  \kapparef
  \widehat{\vartemp} 
  \vectordot{\nabla \widetilde{\Theta}}{\nabla \widetilde{\Theta}}
  \,
  \cvolumee
  ,
\end{equation}
where $\widetilde{\Theta}$ is given by the formula~\eqref{eq:164} written in terms of $\widetilde{\vartemp}$ and $\widehat{\vartemp}$, that is
\begin{equation}
  \label{eq:174}
  \widetilde{\Theta}
  =
  \ln \left( 1 + \frac{\widetilde{\vartemp}}{\widehat{\vartemp}} \right)
  .
\end{equation}
\end{subequations}
Consequently, we see that the time derivative of the proposed functional is nonpositive, and that the derivative vanishes if and only if $\widetilde{\vartemp}=0$ in $\Omega$. This implies that the steady non-equilibrium state $\widehat{\vartemp}$ is unconditionally asymptotically stable.

\subsubsection{Lyapunov function for the original problem}
\label{sec:lyap-funct-orig}
Having shown unconditional asymptotic stability of steady non-equilibrium state $\widehat{\vartemp}$ in the auxiliary heat conduction problem~\eqref{eq:138}, we can go back to the original problem of heat conduction in a body with temperature dependent thermal conductivity and constant specific heat capacity at constant volume, see~\eqref{eq:127}. 

Since the system~\eqref{eq:127} is formally equivalent to~\eqref{eq:138}, we see that the unconditional asymptotic stability of $\widehat{\vartemp}$ is equivalent to the unconditional asymptotic stability of $\widehat{\temp}$. If we want to explicitly construct the Lyapunov functional for~\eqref{eq:127}, and find its time derivative, all that needs to be done is to rescale temperature $\vartemp$ in \eqref{eq:146} and~\eqref{eq:177} using the substitution
\begin{equation}
  \label{eq:175}
  \vartemp = \frac{\tempref}{\alpha} \left( \exponential{\alpha \frac{\temp - \temp_{\reference}}{\temp_{\reference}}} - \exponential{-\alpha}\right),
\end{equation}
see~\eqref{eq:135} and~\eqref{eq:167}. This gives us the Lyapunov functional in terms of the original temperature field $\temp$.


\vfill

\end{document}